\documentclass[12pt]{article}
\usepackage{amsmath}
\usepackage{amssymb}
\usepackage[dvips]{graphics}
\usepackage{epsfig}
\usepackage{calc}
\usepackage{amsfonts}
\usepackage{graphicx}
\usepackage[ansinew]{inputenc}
\usepackage{color}

\catcode`\@=11
\def\marginnote#1{}
\newcount\hour
\newcount\minute
\newtoks\amorpm
\hour=\time\divide\hour by60 \minute=\time{\multiply\hour by60
\global\advance\minute by-\hour}
\edef\standardtime{{\ifnum\hour<12 \global\amorpm={am}%
        \else\global\amorpm={pm}\advance\hour by-12 \fi
        \ifnum\hour=0 \hour=12 \fi
        \number\hour:\ifnum\minute<10 0\fi\number\minute\the\amorpm}}
\edef\militarytime{\number\hour:\ifnum\minute<10
0\fi\number\minute}
\def\draftlabel#1{{\@bsphack\if@filesw {\let\thepage\relax
   \xdef\@gtempa{\write\@auxout{\string
      \newlabel{#1}{{\@currentlabel}{\thepage}}}}}\@gtempa
   \if@nobreak \ifvmode\nobreak\fi\fi\fi\@esphack}
        \gdef\@eqnlabel{#1}}
\def\@eqnlabel{}
\def\@vacuum{}
\def\draftmarginnote#1{\marginpar{\raggedright\scriptsize\tt#1}}
\def\draft{\oddsidemargin -.5truein
        \def\@oddfoot{\sl preliminary draft \hfil
        \rm\thepage\hfil\sl\today\quad\mil922 itarytime}
        \let\@evenfoot\@oddfoot \overfullrule 3pt
        \let\label=\draftlabel
        \let\marginnote=\draftmarginnote
   \def\@eqnnum{(\theequation)\rlap{\kern\marginparsep\tt\@eqnlabel}%
\global\let\@eqnlabel\@vacuum}  }

\def\preprint{\twocolumn\sloppy\flushbottom\parindent 1em
        \leftmargini 2em\leftmarginv .5em\leftmarginvi .5em
        \oddsidemargin -.5in    \evensidemargin -.5in
        \columnsep 15mm \footheight 0pt
        \textwidth 250mmin      \topmargin  -.4in
        \headheight 12pt \topskip .4in
        \textheight 175mm
        \footskip 0pt
        \def\@oddhead{\thepage\hfil\addtocounter{page}{1}\thepage}
        \let\@evenhead\@oddhead \def\@oddfoot{} \def\@evenfoot{} }

\def\titlepage{\@restonecolfalse\if@twocolumn\@restonecoltrue\onecolumn
     \else \newpage \fi \thispagestyle{empty}\c@page\z@
        \def\thefootnote{\fnsymbol{footnote}} }

\def\endtitlepage{\if@restonecol\twocolumn \else  \fi
        \def\thefootnote{\arabic{footnote}} \setcounter{footnote}{0}}

\catcode`@=12 \relax

\def\bea{\begin{array}}
\def\bem{\begin{displaymath}}
\def\beq{\begin{equation}}

\def\Z{\mathop{\bf Z}}

\def\eea{\end{array}}
\def\eem{\end{displaymath}}
\def\eeq{\end{equation}}
\def\half{\frac{1}{2}}

\relax
\def\be{\begin{equation}}
\def\ee{\end{equation}}
\def\ba{\begin{eqnarray}}
\def\ea{\end{eqnarray}}

\def\del{\partial}
\def\d{{\rm d}}

\def\r{\rho}
\def\a{\alpha}
\def\b{\beta}

\def\e{\epsilon}

\def\o{\omega}

\def\l{\lambda}
\def\L{\Lambda}

\def\cN{{\cal N}}

\def\cF{{\cal F}}
\def\cR{{\cal R}}
\def\cRh{\widehat{\cal R}}
\def\St{{\widetilde S}}

\def\cFh{\widehat\cF}
\def\Sh{\hat S}

\def\Wh{\widehat W}
\def\Weff{W_{\rm eff}}
\def\Wheff{\Wh_{\rm eff}}
\def\yh{\widehat y}

\def\IR{\relax{\rm I\kern-.18em R}}

\def\inv{^{\raise.15ex\hbox{${\scriptscriptstyle -}$}\kern-.05em 1}}

\def\be{\begin{equation}}
\def\ee{\end{equation}}
\def\ba{\begin{eqnarray}}
\def\ea{\end{eqnarray}}

\def\tr{\,{\rm tr}\,}

\def\hN{\hat{N}}

\def\wh{\widehat}
\def\wt{\widetilde}
\def\yt{\wt y}
\def\Lh{{\wh \L}}

\relax

\renewcommand{\theequation}{\thesection.\arabic{equation}}
\begin{document}
\topmargin-2.4cm
%
%
%
%
\begin{titlepage}
\begin{flushright}
LPTENS-06/02\\
LMU-ASC 06/06\\
hep-th/0601231
\\
revised May 15, 2006
\end{flushright}
\vskip 2.5cm

\begin{center}{\Large\bf Relating prepotentials and quantum vacua of } \\
\vspace{2mm} {\Large \bf $\cN=1$  gauge theories }\\
\vspace{2mm} {\Large \bf  with different tree-level superpotentials}
\vskip 1.5cm {\bf Adel Bilal$^{1}$ and Steffen
Metzger$^{1,2}$}

\vskip.3cm $^1$ Laboratoire de Physique Th\'eorique,
\'Ecole Normale Sup\'erieure - CNRS\\
24 rue Lhomond, 75231 Paris Cedex 05, France

\vskip.3cm $^2$ Arnold-Sommerfeld-Center for Theoretical
Physics,\\
Department f\"ur Physik, Ludwig-Maximilians-Universit\"at\\
Theresienstr. 37, 80333 Munich, Germany\\

\vskip.3cm {\small e-mail: {\tt adel.bilal@lpt.ens.fr,
steffen.metzger@physik.uni-muenchen.de}}
\end{center}
\vskip .5cm

\begin{center}
{\bf Abstract}
\end{center}
\begin{quote}
We consider ${\cal N}=1$ supersymmetric  $U(N)$ gauge theories
with $\Z_k$ symmetric tree-level superpotentials $W(\Phi)$ for an adjoint 
chiral multiplet. We show that (for $2N/k\in\Z$)
this $\Z_k$ symmetry survives in the quantum effective theory as a
corresponding symmetry of the effective superpotential $\Weff(S_i)$
under permutations of the $S_i$. For  $W(x)=\Wh(\xi)$, $\xi=x^k$, this 
allows us to express the prepotential $\cF_0$ and effective 
superpotential $\Weff$ on certain submanifolds of the moduli 
space in terms of an $\cFh_0$ and $\Wheff$ of a different theory 
with tree-level superpotential $\Wh$. In particular, 
if the $\Z_k$ symmetric polynomial $W(x)$ is of degree $2k$, then $\Wh$ is 
gaussian and we obtain very explicit formulae for $\cF_0$ and $\Weff$. 
Moreover, in this case, every vacuum of the effective 
Veneziano-Yankielowicz superpotential $\Wheff$ is shown to give 
rise to a vacuum of $\Weff$. Somewhat surprisingly, at the level 
of the prepotential $\cF_0(S_i)$ the permutation symmetry only 
holds for $k=2$, while it is anomalous for $k\ge 3$ due to subtleties 
related to the non-compact period integrals. Some of these results are 
also extended to general polynomial relations $\xi=h(x)$ between 
the tree-level superpotentials.
\end{quote}


\end{titlepage}
\setcounter{footnote}{0} \setcounter{page}{0}
\setlength{\baselineskip}{.6cm}
\newpage
%
%
\newtheorem{proposition}{Proposition}[section]
\newtheorem{theorem}[proposition]{Theorem}
\newtheorem{definition}[proposition]{Definition}
\newtheorem{conjecture}[proposition]{Conjecture}

\section{Introduction\label{Intro}}
\setcounter{equation}{0}

Understanding the vacuum structure of strongly coupled gauge theories 
remains an important challenge. Considerable progress has been made 
over the last years within the framework of ${\cal N}=1$ supersymmetric 
gauge theories in  computing the exact quantum-effective 
superpotential $\Weff$ \cite{KKV}-\cite{DV}. This involved geometric 
engineering 
of the gauge theory within string theory \cite{KKV,KKLM} and computation of 
the topological string amplitudes \cite{GT} on local Calabi-Yau 
manifolds, geometric transitions and large $N$ dualities \cite{VAFA,CIV}, 
and 
culminated in the realisation that the non-perturbative effective 
superpotential $\Weff$ can be directly obtained from an appropriate 
holomorphic matrix model in the planar limit \cite{DV}. Later on, 
these results were also obtained within field theory \cite{CDSW}.

While this program has been carried out for various gauge groups and 
matter contents, here we will 
only consider the simplest case of $\cN=1$ supersymmetric Yang-Mills 
theory with 
$U(N)$ gauge group coupled to an adjoint chiral multiplet $\Phi$ with a 
tree-level
superpotential $W(\Phi)$. If $W(\Phi)$ is of order $n+1$ having $n$ 
non-degenerate 
critical points, a general vacuum breaks the gauge group
to $\prod_{i=1}^n U(N_i)$ with $\sum_{i=1}^n N_i = N$. 
This gauge theory can be obtained from IIB string theory on a specific 
local Calabi-Yau manifold \cite{KKLM} which can be taken through a 
geometric transition.
The geometry of the local Calabi-Yau manifold after the geometric transition
is directly determined by $W(x)$ together with $n$ deformation parameters 
(complex structure moduli) which are encoded in the coefficients 
of a polynomial $f(x)$ of order $n-1$. This geometry is closely 
linked to the hyperelliptic Riemann surface
\be\label{Riemann}
y^2=W'(x)^2 +f(x)\ ,
\ee
which also appears in the planar limit of the holomorphic matrix model 
with action $\tr W(M)$. 
In the gauge theory, for each $U(N_i)$-factor the $SU(N_i)$ confines, 
and the low-energy dynamics is described by $n$ $U(1)$-vector multiplets 
together with the chiral ``glueball'' superfields
$S_i\sim {\rm tr}_{SU(N_i)} W_\a W^\a$. This is an ${\cal N}=2$ theory 
softly broken to ${\cal N}=1$ by some effective superpotential 
$\Weff(S_i)$ which one needs to compute. Also, the $U(1)^n$ couplings 
are given 
as second derivatives of a prepotential $\cF_0(S_i)$. 
The effective superpotential and the prepotential are essentially 
given in terms of period integrals  \cite{CIV,DV,BM} on the Riemann 
surface (\ref{Riemann}). They 
can be divided into $A$ and $B$ periods, with the $A_i$ periods 
giving (the lowest components of) the chiral superfields $S_i$ while the 
$B_i$ periods are given as $\del \cF_0/\del S_i$, up to some divergent 
terms \cite{BM}, revealing the rigid special geometry. The function 
$\cF_0(S_i)$ is related to the genus zero free energy of the topological 
string on the local Calabi-Yau manifold and can also be identified with 
the matrix model planar free 
energy with fixed filling fractions. We will refer to it as the prepotential.

The original $U(N)$ super Yang-Mills theory (in the absence of the tree-level
superpotential $W(\Phi)$) has various global $U(1)$ symmetries acting on 
$\Phi$ that are broken to discrete subgroups by the usual anomaly.
However, a non-vanishing generic $W(\Phi)$ completely breaks even
these remaining anomaly-free discrete symmetries. In this note, we are 
interested in the case where the tree-level superpotential has certain 
discrete symmetries and preserves a corresponding anomaly-free discrete
subgroup. Our aim is to explore as much as possible the implications of
these symmetries on the effective superpotential $\Weff$, as well as on
the prepotential $\cF_0$.

Suppose that $W(\Phi)$ has a $\Z_k$ symmetry, i.e. it is a sum of terms 
of the form $\tr\Phi^k,\ \tr\Phi^{2k}$, etc. Consider the $U(1)$ symmetry
$\Phi\to e^{i\a}\Phi$, with the superspace 
coordinate $\theta$ and the gauge multiplet $W_\b$ unaffected. 
(This is not an $R$-symmetry.) Due to the anomaly\footnote{
The anomalous transformation of the fermion measure gives, as usual, 
an extra factor 
$\exp\left( i{\a \over 8\pi^2}\int {\rm tr}_{\rm ad} F\wedge F\right)
=\exp\left( i\a\,2N\,\nu\right)$ where $\nu$ is the instanton number. 
}
this $U(1)$ is broken down to ${\bf Z}_{2N}$, i.e. 
$\a=2\pi {r\over 2N}, \ r=1,\ldots 2N$. For a generic term in the 
superpotential we have $\tr\Phi^l\to e^{i l \a} \tr \Phi^l 
= e^{2\pi i {l r\over 2N} } \tr \Phi^l$ and a general superpotential 
(containing at least two terms  $\tr\Phi^l$ and  $\tr\Phi^{l'}$ with $l$ 
and $l'$ having no common divisor)
breaks the $\Z_{2N}$ completely. However, if $W(\Phi)$ has a ${\bf Z}_k$ 
symmetry and $k$ divides $2N$, say ${2N\over k}=s\in \Z$, then for all terms 
in $W(\Phi)$ we have $l=k\, p,\ p\in\Z$ and 
$\tr \Phi^{kp} \to  e^{2\pi i p{r\over s} } \tr \Phi^{kp}$ which
is invariant for $r=s, 2s, \ldots ks=2N$, i.e. the superpotential 
indeed preserves a $\Z_k$ subgroup of the anomaly-free $\Z_{2N}$.

This non-anomalous ${\bf Z}_k$ acts on $\Phi$ as $\Phi\to e^{2\pi i q/k}\Phi$
and, in particular, permutes among themselves the solutions of $W'(\Phi)=0$, 
which are the classical vacua. Hence, it must also permute the eigenvalues 
sitting at 
(or close to) the critical points accordingly. In the effective theory
which is described by the $S_i$ one thus expects that  permuting
the corresponding values of the $S_i$ is a symmetry. This is indeed the 
case, as we will show in this note.

In the simplest case $k=2$, $W(x)$ is an even 
function of $x$, and then we may write $W(x)=\half \Wh(x^2)$.
More generally, for superpotentials with $\Z_k$-symmetry 
generated by $x\to e^{2\pi i/k} x$ we write
\be\label{xk}
W(x)={1\over k} \Wh(\xi) \ , \quad \xi=x^k \ .
\ee
Of course, if $W$ is of order $n+1$ and $\Wh$ of order $m+1$, we must have
\be\label{orders}
n+1=k(m+1) \ .
\ee
The simplest non-trivial example is $m=1$, $k=2$ where $\Wh$ is a 
quadratic (gaussian) superpotential and $W$ a quartic one. 

Our basic observation is that (\ref{xk}) induces a map between the two
Riemann surfaces $\cR$  given by $y^2=W'(x)^2+f(x)$ 
and $\cRh$ given by $\yh^2=\Wh'(\xi)^2+\hat f(\xi)$. 
We will exploit this to compute and relate the corresponding period 
integrals, prepotentials and effective superpotentials, thus generalising 
the simple geometric map (\ref{xk}) to a map between two different 
quantum gauge theories which we will call I and II. We will systematically 
exploit the $\Z_k$ symmetry to show that there is a corresponding $\Z_k$ 
symmetry of the effective theory described by the $S_i$ (modulo a 
mild anomaly discussed below). In a loose way, 
one might think of theory II as being the ``quotient'' of theory I 
by this $\Z_k$.

For general $m$ and $k$, the 
corresponding super Yang-Mills theories have breaking patterns\footnote{
Note that 
relations between different theories having the {\it same} tree-level 
superpotential but corresponding to different gauge symmetry breaking 
patterns were examined e.g. in \cite{CSW}. In this case $n=m$, 
which is quite different from the relations we are considering.
}
\ba\label{breaking}
{\rm I}\ : \quad 
&&U(N)\to\left(\prod_{l=1}^m \prod_{r=1}^k U(N_{l,r})\right) 
\times 
\prod_{s=1}^{k-1} U(N_{0,s}) 
\quad {\rm and} \\
{\rm II}\ : \quad &&U(\hN)\to \prod_{j=1}^m U(\hN_j) \ .
\ea
Of course, in general, theory I depends on the $S_i$, with
$i=1,\ldots n=km+k-1$ while theory II only depends on much less fields 
$\Sh_j$, $j=1,\ldots m$. We will be able to relate the theories if 
precisely $k-1$ of the $S_i$ (called $S_{0,s}$, $s=1,\ldots k-1$) vanish, 
and if for the remaining $S_i\equiv S_{l,r}$, with $l=1,\ldots m$ and 
$r=1,\ldots k$, we have $S_{l,r} ={\Sh_l\over k}$. In particular, we will
show that we can relate the prepotentials $\cF_0$ and $\cFh_0$:
\be\label{F0rel}
\cF_0\Big( S_{0,s}=0;\, S_{l,r}={\Sh_l\over k}\Big) 
= {1\over k} \cFh_0(\Sh_l) \ .
\ee
Moreover, we can also relate the
{\it effective} superpotentials $\Weff$ and $\Wheff$
provided $N_{l,r}={\hN_l\over k} \, ,\ N_{0,s}=0$ 
(we always take $l=1,\ldots m$ and $r=1,\ldots k$, 
as well as $s=1,\ldots k-1$). 
For these choices of $N_i$ we will show that
\be\label{Weffrel}
\Weff\Big(S_{0,s}=0;\, S_{l,r}={\Sh_l\over k}\Big) 
= {1\over k} \Wheff(\Sh_l) \ .
\ee
Note that these choices of $N_{l,r}$ and $N_{0,s}$ imply that
the $\hN_l$ are multiples of $k$ and hence that $N$ and a fortiori
$2N$ is a multiple of $k$. This was our condition for $\Z_k$ to be 
an anomaly-free symmetry of the $U(N)$ super Yang-Mills theory!

We also want to determine vacua of the quantum theory, and then
one needs to find extrema of $\Weff$ with respect to {\it independent} 
variations of all $S_{l,r}$.  (The $S_{0,s}$ are not varied and 
remain zero if $N_{0,s}=0$).
For general $S_{l,r}$ we are able to show 
that the $\Z_k$-symmetry of $W(x)$ implies a corresponding quantum 
symmetry\footnote{
It is interesting to note that, somewhat similarly, the interplay 
of physical and ``geometric'' $\Z_l$ symmetries has often been useful. 
In ${\cal N}=2$ super Yang-Mills theories discrete subgroups of $U(1)_R$ 
symmetries give rise to ``geometric'' $\Z_l$ symmetries on the moduli 
space of vacua which was one of the key ingredients in the determination
of the spectra of stable BPS states in \cite{BF}. The arguments in 
the recent work \cite{FF} to relate discrete gauge invariance and
the analytic structure of $\langle\det(z-\Phi)\rangle$ also 
uses permutations between eigenvalues located on different cuts 
and is somewhat reminiscent to the arguments we use in the present note.
} 
of $\Weff$ under cyclic permutations 
$S_{l,r}\to S_{l,r+1},\ S_{l,k}\to S_{l,1}$. 
For the special cases of $m=1$ (and arbitrary $k$),
this symmetry, in turn, can be exploited
to show that $\Weff$ has indeed an extremum at 
$S_{1,1}=\ldots S_{1,k}=S^*$ with respect to independent variations 
of all $S_{1,r}$, with $S^*$ determined by the minimum 
of $\Wheff(S)$, i.e. of the Veneziano-Yankielowicz effective superpotential.
One would expect that, similarly, the prepotential 
$\cF_0$ is symmetric under cyclic 
permutations of  unequal $S_{l,r}$. While this is true for $k=2$, 
it is no longer the case for $k\ge 3$ due to subtleties related to 
the common choice of cutoff for all $B_{l,r}$ cycles which breaks 
the $\Z_k$ symmetry. This is very much 
like an anomaly. Of course, there is nothing wrong with such an 
anomaly since it concerns a global discrete symmetry. Furthermore,
the physical quantity in the gauge theory, $\Weff$, is not 
affected by this anomaly. It would be interesting to explore whether
 there are physical observables beyond the gauge theory that are 
sensitive to this anomaly.

Note that for $m=1$ (gaussian), $\cFh_0$ and $\Wheff$ are explicitly 
known functions. For $m=2$, the Riemann surface is a punctured torus, 
and $\cFh_0$ and $\Wheff$ can still, in principle, be expressed through 
various combinations of complete and incomplete elliptic functions. For 
$m\ge 3$, {\it in general}, no explicit expressions in terms of special 
functions are known. Our mappings between theories constitute precisely 
the exceptions where, for $n\ge 3$ explicit expressions can nevertheless 
be obtained.

This paper is organised as follows. In Sect. 2, we briefly review the 
formalism we will use and introduce some notation. For a detailed 
review we refer to \cite{SM}. We also recall some subtleties related 
to the definition of the non-compact (relative) $B$ cycles and the 
evaluation of the corresponding period integrals (see \cite{BM} for details). 
In particular, we give a useful formula expressing the prepotential 
$\cF_0$ solely in terms of integrals over (relative) cycles on the 
Riemann surface. In Sect. 3, we establish 
the various relations between theories I and II. We start (Sect. 3.1) 
with the simplest case of an even quartic tree-level superpotential 
$W(x)$ (theory I) which is mapped via $\xi=x^2$ to a gaussian tree-level 
superpotential $\Wh(\xi)$ (theory II). This warm-up exercise already 
contains all the ideas but little technical complications. In particular, 
we relate $\cF_0$ to $\cFh_0$, $\Weff$ to $\Wheff$ (which is the 
Veneziano-Yankielowicz superpotential) for $S_1=S_2=\Sh/2\equiv t/2$, 
prove the symmetries of $\cF_0$ and $\Weff$ under exchange of 
unequal $S_1$ and $S_2$,
and show that each vacuum of $\Wheff$ (theory II) gives rise to a 
vacuum of $\Weff$ (theory I).\, Sect. 3.2 deals with a general even 
$W(x)$ which, by $\xi=x^2$, can be mapped to a (general) $\Wh(\xi)$. 
Here we can still relate 
$\cF_0$ to $\cFh_0$ and $\Weff$ to $\Wheff$ and 
prove the symmetry properties, but, in general, we do not 
know the {\it explicit} expressions of $\cFh_0$ or $\Wheff$. In 
Sect. 3.3, we study tree-level superpotentials $W(x)$ of order 
$2k$ having a $\Z_k$-symmetry, so that one can use $\xi=x^k$ to map 
them to a gaussian $\Wh(\xi)$. Although conceptually this is very similar 
to the case studied in Sect. 3.1, there are various technical subtleties, 
related to the precise definition of the $B_i$ cycles, which have to be 
addressed. In the end, we can still relate $\cF_0$ to $\cFh_0$ and 
$\Weff$ to $\Wheff$ by (\ref{F0rel}) and (\ref{Weffrel}), and show, 
moreover, that 
for each vacuum of the Veneziano-Yankielowicz superpotential $\Wheff$ we 
get a vacuum of $\Weff$. We discuss in detail how the permutation anomaly of
$\cF_0$ arises and why the symmetry is restored for $\Weff$.
Sect. 3.4 discusses a general $\Z_k$-symmetric 
$W(x)$ of degree $k(m+1)$. Again, relations (\ref{F0rel}) and 
(\ref{Weffrel}) hold but, in general, we lack explicit expressions 
for $\cFh_0$ 
or $\Wheff$. Finally, in Sect. 3.5, we comment on general maps 
$\xi=h(x)$ and $W(x)={1\over k} \Wh(\xi)$. Equations  (\ref{F0rel}) 
and (\ref{Weffrel}) continue to be true, but without the $\Z_k$-symmetry 
we were not able to determine any vacuum of $\Weff$ from the vacua of $\Wheff$,
even for $m=1$. To conclude, in Sect. 4, we present a table summarising 
our results for the various cases.

\section{The tools\label{tools}}
\setcounter{equation}{0}

As first conjectured in \cite{VAFA,CIV}, and motivated by the geometric 
transition between local Calabi-Yau manifolds \cite{GT}, in general 
the effective superpotential  $\Weff(S_i)$ is given by
\be\label{weff}
\Weff(S_i)=-\sum_{i=1}^n \left[ N_i {\del \cF_0\over\del S_i}(S_j)
-\a_i(\L,N_j) S_i \right] \ .
\ee
The $S_i$ are the chiral superfields 
whose lowest components are the gaugino bilinears in the $U(N_i)$-factors.
The $N_i$ can be interpreted, in IIB string theory, 
as the numbers of $D5$-branes wrapping 
the $i^{\rm th}$ two-cycle in the Calabi-Yau geometry before the 
geometric transition. After the geometric transition, the $N_i$ are 
given by the integrals of the 3-form field strength 
$H=H_{\rm RR}+\tau H_{\rm NS}$ over the compact 3-cycles which have 
replaced the 2-cycles. The $\a_i$, on the other hand, are given in 
terms of the integrals of the same 3-form over the non-compact 3-cycles. 
They can be viewed as functions of the $N_i$ and the renormalised
$U(N)$ gauge-coupling constant $\tau$, or equivalently the 
physical scale $\L$. In particular, they are independent of the $S_i$ 
which play the role of complex structure moduli. The precise form of 
the $\a_i$ does not concern us here. We will only need the following 
symmetry property: if we permute the $N_j$, then the $\a_i$ are 
permuted accordingly.\footnote{
This is certainly true for the Calabi-Yau geometries resulting from 
a $W(x)$ with a $\Z_k$-symmetry, as studied below. However, we are 
not aware of a proof of this property and we will take it as a hypothesis.
}
In particular, if all $N_j$ are equal, then all $\a_i$ are equal, too.

\begin{figure}[h]
\centering
\includegraphics[width=1.0\textwidth]{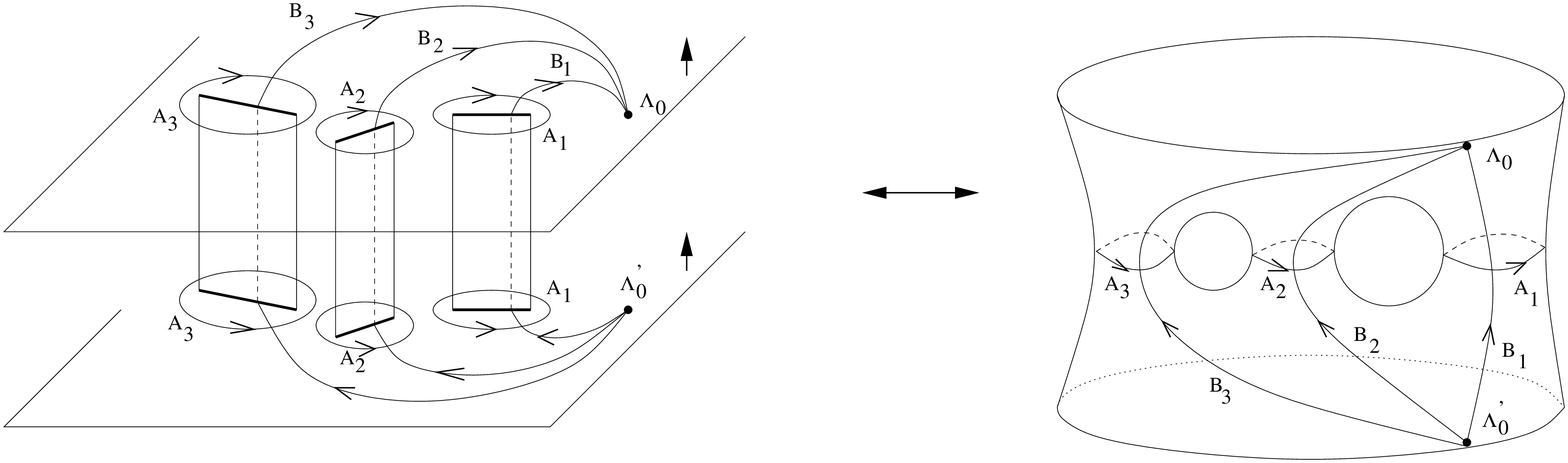}\\
\caption[]{A symplectic choice of compact $A$- and non-compact
$B$ cycles for $n=3$. Note that the orientation of the two planes
on the left-hand side is chosen such that both normal vectors
point to the top. This is why the orientation of the $A$ cycles is
different on the two planes. To go from the representation of the
Riemann surface on the left to the one on the right one has to
flip the upper plane.} \label{x3-0}
\end{figure}

The prepotential $\cF_0$ 
can be obtained from the genus $(n-1)$ hyperelliptic Riemann surface 
given by (\ref{Riemann}).
Here $f(x)$ is a polynomial of order $(n-1)$ depending on $n$ coefficients 
in one-to-one correspondence with the $S_i$ given by
(we use $S_i$ interchangeably 
to denote the superfield or its lowest component) 
\be\label{Siint}
S_i={1\over 4\pi i}\int_{A_i} y(x)\d x \ ,
\ee
where the $A_i$ cycle encircles clockwise the $i^{\rm th}$ cut on the 
upper sheet,\footnote{
Note that the way we number the cuts and corresponding cycles 
is different from \cite{BM}. This will simplify notations later on.
} 
see Fig.\ \ref{x3-0}.
The  prepotential $\cF_0$ or rather 
${\del \cF_0\over\del S_i}$ then is given in terms of integrals 
over {\it non}-compact dual cycles $B_i$. This involves the 
introduction of a cut-off $\L_0$ and, as carefully discussed in 
\cite{BM}, the {\it cut-off independent} result is
\be\label{Biint}
{\del \cF_0\over\del S_i} = \half \int_{B_i} y(x)\d x
-W(\L_0) + \left( \sum_j S_j\right) \log\L_0^2 +o\left({1\over \L_0}\right)
\ ,
\ee
where now $B_i$ runs from $\L_0'$ on the lower sheet through the 
$i^{\rm th}$ cut to $\L_0$ on the upper sheet.

\vskip-3.mm
The Riemann surface (\ref{Riemann}) also appears in the planar limit of 
the corresponding 
matrix model with {\it potential} $W(x)$, as has been known 
for a long time \cite{BIPZ}. The physical reason why computing the effective 
superpotential in ${\cal N}=1$ super Yang-Mills theory reduces to a  
matrix model, actually a  holomorphic matrix model,
was first discovered in \cite{DV}. The holomorphic matrix model
involves integration of the eigenvalues of $\bar N\times \bar N$ matrices
over a specific path $\l(s)$ 
in the complex plane\footnote{
By holomorphicity, the path $\l(s)$ can be chosen arbitrarily except 
that its asymptotics must be such that 
$\left\vert \exp\left(-{1\over g_s} W(\l(s))\right)\right\vert \to 0$. 
However, as discussed 
in \cite{BM}, to get a consistent saddle-point approximation (which 
is actually what one means by the ``planar'' limit), it must 
be such that it goes through the $\l_i^*$ that constitute the solution of 
the saddle-point equations. This implies that all the cuts of $y$ must 
lie on $\l(s)$.
} 
as discussed in detail in \cite{LAZ,BM}.
In this context, $\cF_0$ can be identified with the matrix model planar
free energy 
\be\label{F0}
\cF_0=
t^2\,\mathcal{P}\int\d s\,\d s'\, \log(\l(s)-\l(s'))\r_0(s)\r_0(s')
-t\int\d s\ W(\l(s))\r_0(s)\ ,
\ee
where $\r_0(s)\ge 0$ is the density of eigenvalues (with respect to 
the real parameter $s$ along the path $\l(s)$) and it is given by
\be\label{rhodef}
\r_0(s):=\dot{\l}(s)\lim_{\e\rightarrow 0}{1\over4\pi i t}
[y_+(\l(s)+i\e\dot{\l}(s))-y_+(\l(s)-i\e\dot{\l}(s))]\ ,
\ee
i.e. by the discontinuity of $y_+$ ($y_+$ denotes the value of 
$y$ on the upper sheet) 
across its branch cuts.
The parameter $t$ is the 't Hooft coupling $t=g_s \bar N$ where  
${1\over g_s}$ is the coefficient in 
front of $W(M)={1\over n+1}\tr M^{n+1}+\ldots$ in the matrix model action. 
It is easy to check (see e.g. \cite{BM}) that this $\r_0(s)$ is correctly 
normalised provided 
one identifies the leading coefficient of the polynomial $f(x)$ in 
(\ref{Riemann}) with $-4t$. Note also that $\r_0$ depends on the 
$S_i$ which are the moduli of the Riemann surface (\ref{Riemann}), 
and that $\sum_{i=1}^n S_i=t$.

In much of the matrix model literature, the parameter $t$ is fixed to 
some convenient value. Here, however, it is crucial to keep  $t$ arbitrary,
and hence the $S_i$ unconstrained, so that we really have $n$ independent 
moduli\footnote{
Note that this is different from naive expectations for a compact 
genus $g=n-1$ Riemann surface. In particular, for $n=1$, the sphere 
has no (complex structure) modulus at all. However, we are dealing 
with non-compact 
surfaces and, for $n=1$, we actually have a sphere with marked 
points $\L_0$ and $\L_0'$, or actually a sphere with two disks 
around the north and south pole deleted. The ratio $t/\L_0$ measures 
the size of these holes.
}
and $\cF_0$ is a function of all $S_i$. In practice, eq. (\ref{F0})
is not always convenient to actually compute $\cF_0$. In \cite{BM}, we derived
an alternative formula which more directly uses the period integrals 
of (\ref{Siint}) and (\ref{Biint}), namely\footnote{
Eq. (3.64) of ref. \cite{BM} actually uses a different basis of 
cycles and is written in a slightly different but equivalent form.
A similar formula also appeared in \cite{CSW}.}
\be\label{homrel}
\cF_0(S_i) = \half \sum_{i=1}^n S_i {\del \cF_0\over \del S_i} 
-{t\over 2} \int\d s\, \r_0(s) W(\l(s)) \ .
\ee
The last integral reduces to a sum over integrals over the cuts. Using 
(\ref{rhodef}) it is easily rewritten as a sum of contour integrals and we get
\be\label{homrelbis}
\cF_0(S_i) = \half \sum_{i=1}^n \left[ S_i {\del \cF_0\over \del S_i} 
-{1\over 4\pi i} \int_{A_i} W(x) y(x) \d x \right] \ .
\ee
In view of (\ref{Siint}) and (\ref{Biint}), this expresses $\cF_0$ 
entirely in terms of integrals over the $A$ and $B$ cycles of the 
Riemann surface.

In \cite{BM} we studied how $\cF_0$ changes under symplectic changes of 
basis of $A$ and $B$ cycles. A particularly simple symplectic change is
\be\label{symplecticchange}
B_i\to B_i +\sum_j n_{ij} A_j \ , \quad n_{ij}=n_{ji}\in\Z \ .
\ee
It follows from (\ref{Biint}) and (\ref{Siint}) that 
${\del \cF_0\over \del S_i} \to 
{\del \cF_0\over \del S_i} + 2\pi i\sum_j n_{ij} S_j$ and hence 
from (\ref{homrelbis}) that \cite{BM}
\be\label{F0change}
\cF_0\to \cF_0+i\pi \sum_{i,j} S_i n_{ij} S_j \ .
\ee
It is quite interesting to note that equation (\ref{F0}) gives $\cF_0$
directly in terms of the eigenvalue density $\r_0$ and seems not to be 
concerned about how one chooses the exact form of the $B_i$ cycles. 
However, it involves a double integral with a logarithm and, to be 
precise, one has to choose the branches of the logarithm. Choosing 
different branches results in adding to $\cF_0$ a quadratic form 
$i \pi \sum_{i,j} S_i n_{ij}  S_j$ with even integers $n_{ij}=n_{ji}$. This is 
in agreement with (\ref{F0change}), except that only even $n_{ij}$ 
appear. Indeed, the integrals in (\ref{F0}) are defined on the cut 
$x$-plane and changing the branches of the logarithm corresponds to 
a performing a monodromy where the cut ${\cal C}_i$ goes once around the cut 
${\cal C}_j$. Under such a monodromy one has $B_i\to B_i\pm 2 A_j$ 
and $B_j\to B_j\pm 2 A_i$, necessarily with an even $n_{ij}$.

It will be useful to recall the results for the simplest case $n=1$:
\be\label{gaussianpot}
n=1 \ : \quad W(x)=\half (x-a)^2 + w_0 \ , \quad f(x)=-4t
\ee
Then we have a single pair of $A$ and $B$ cycles. From eqs. (\ref{Siint}), 
(\ref{Biint})  and (\ref{homrel}) one gets
\ba\label{n=1int}
n=1 \ : \quad && S=t \ , \qquad {\del \cF_0\over \del t}= t\log t - t - w_0 \ ,
\nonumber\\
&& \cF_0(t)={t^2\over 2} \log t -{3\over 4} t^2 - t\, w_0 \ .
\ea
Note that $\cF_0$ is real for $t>0$ (and real $w_0$). 
A different choice of $B$ cycle
as in (\ref{symplecticchange}) would have resulted in a complex $\cF_0$.

\section{Relating different theories\label{relating}}
\setcounter{equation}{0}

Now we are ready to relate the free energies, resp. prepotentials, $\cF_0$ 
and effective superpotentials $\Weff$ of different matrix 
models, resp. different gauge theories.

\subsection{Quartic even superpotential}

As a warm-up exercise, we consider the case $n=3$ with a quartic 
superpotential which we require to be an even ($\Z_2$ symmetric)
function of $x$:
\be\label{quartic}
W(x)={1\over 4} x^4 -{a\over 2} x^2 + b \ .
\ee
If we let $\xi=x^2$ and $w_0=2 b-{a^2\over 2}$ we have
\be\label{wqgrel}
\xi=x^2 \ \ : \qquad
W(x)=\half \Wh(\xi) \ , \quad \Wh(\xi)=\half (\xi-a)^2 +w_0 \ .
\ee
The quartic superpotential $W(x)$ has three critical points at the 
zeros of $W'(x)=x^3-a x$, i.e. $\sqrt{a},\ -\sqrt{a},\ 0$. 
Of course, the $Z_2$ symmetry exchanges $\sqrt{a}$ and $-\sqrt{a}$ 
and leaves $0$ invariant. Generically, 
this leads to three cuts\footnote{
Again, the way we label the cuts is unimportant and only of 
notational convenience.
}
${\cal C}_1, {\cal C}_2$ and ${\cal C}_0$ 
of different size, parametrised by three different $S_1, S_2$ and 
$S_0$ or, equivalently, by the three coefficients of 
$f(x)=-4t x^2 +f_1 x + f_0$. However, if we choose $f_1=f_2=0$, we 
not only respect the $\Z_2$ symmetry, but
\be\label{yqgrel}
y^2(x)=W'(x)^2+f(x) = x^2 \left[ (x^2-a)^2-4t\right] \equiv x^2\ \yh^2(x^2)
\ee
is such that the critical point at $x=0$ does not open to a branch cut, 
while the 
cuts that develop at $x=\pm\sqrt{a}$ have the same size. They
both correspond to the single cut in the 
$\xi=x^2$-plane from $a-2\sqrt{t}$ to $a+2\sqrt{t}$. 
We call $\cR$, resp. $\cRh$, the Riemann surfaces whose sheets are the 
upper and lower $x$-planes, resp. $\xi$-planes. From the preceeding 
construction we see that the 
$t$-moduli  of both Riemann surfaces, $t$ and $\widehat t$, coincide:
\be\label{equalt}
\widehat t = t\ .
\ee
It also follows that 
$y(x)$ is an {\it odd} function of $x$ on both sheets and that we have
\be\label{yyhat}
y(x)\,\d x=+ \half \yh(\xi)\,\d\xi \ .
\ee
As a consequence,
\be\label{sqsg}
\int_{A_1} y(x)\,\d x =\int_{A_2} y(x)\,\d x  =\half \int_A \yh(\xi)\, \d\xi
\quad \Rightarrow\quad S_1=S_2={t\over 2} \ ,
\ee
since both cycles $A_1$ and $A_2$ of $\cR$ are mapped to the 
$A$ cycle of $\cRh$, see Fig.\ 2. Obviously also, 
$S_0=0$. 
\begin{figure}[h]
\centering
\includegraphics[width=0.8\textwidth]{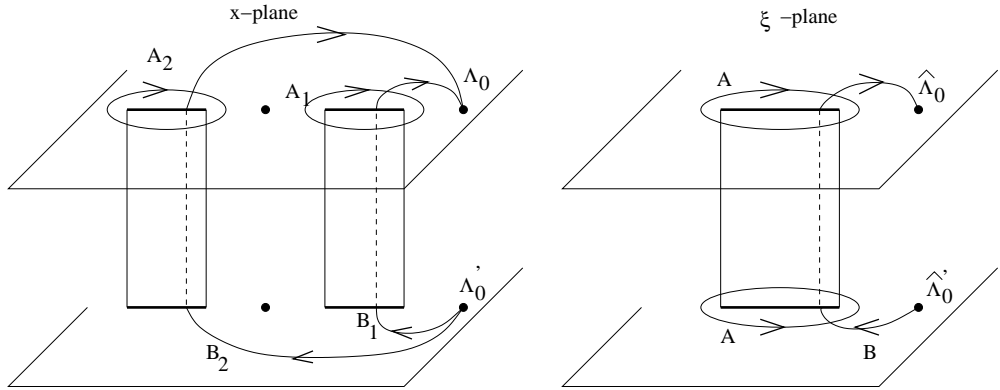}\\
\caption[]{Shown are the cuts and cycles of $\cR$ for the quartic 
superpotential (on the left) and of $\cRh$ for the corresponding quadratic 
superpotential (on the right).} \label{quarticfig}
\end{figure}
\vskip-2.mm

\begin{figure}[h]
\centering
\includegraphics[width=0.9\textwidth]{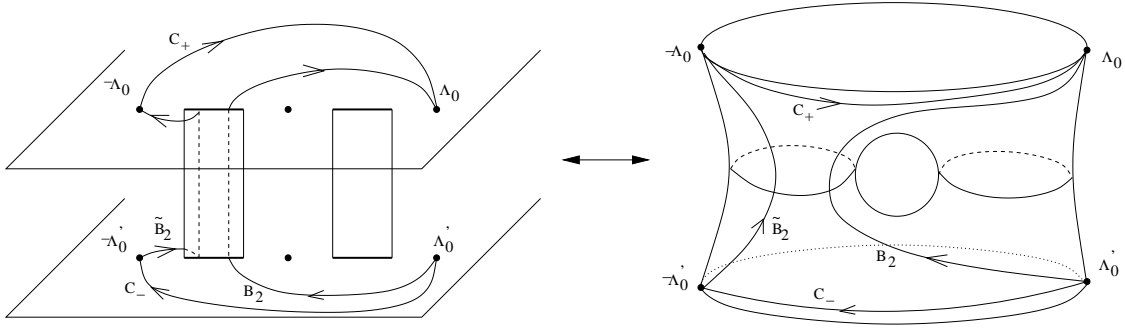}\\
\caption[]{Shown is the decomposition of the $B_2$ cycle into $C_-$, 
$\widetilde B_2$ and $C_+$ for the two different representations 
of the Riemann surface $\cR$.} \label{B1cycle}
\vskip-3.mm
\end{figure}

For the non-compact $B_1$ and $B_2$ cycles one has to be
more careful. Obviously, for the $B$ cycle we choose start and end points
$\widehat \L_0'=(\L_0')^2$ on the lower sheet 
and $\widehat \L_0=(\L_0)^2$ on the upper sheet. Then
the $B_1$ cycle is indeed mapped to the $B$ cycle. However, this is not 
immediately obvious for the $B_2$ cycle. Instead we have
\be\label{B1integrals}
\int_{B_2} y(x)\,\d x =
\int_{C_-} y(x)\,\d x +
\int_{\widetilde B_2} y(x)\,\d x +
\int_{C_+} y(x)\,\d x \ ,
\ee
where $C_-$ goes from $\L_0'$ to $-\L_0'$ on the lower sheet, 
$\widetilde B_2$ from $-\L_0'$ through the cut to $-\L_0$ on the 
upper sheet, and $C_+$ from  $-\L_0$ to $\L_0$, as indicated in 
Fig.\ \ref{B1cycle}.

Now, $C_-$ is mapped to $-A$ in the $\xi$-plane, $C_+$ to $A$ and
$\widetilde B_2$ to $B$. As a result, the two integrals over $C_+$ 
and $C_-$ cancel\footnote{
It is easy to check this statement directly by taking $C_\pm$ to be 
large semicircles so that one can use the asymptotic form 
$y(x)=\pm \left( W'(x) -{2t\over x}\right)$ with the $\pm$ sign 
on the upper/lower plane.
} 
and the integral over $B_2$ equals the integral
over $\widetilde B_2$. Hence
\be\label{B1rel}
\int_{B_2} y(x)\,\d x =\int_{B_1} y(x)\,\d x
=\half \int_{B} \yh(\xi)\,\d \xi \ .
\ee
Using these relations in eq. (\ref{Biint}), together with 
$W(\L_0)=\half \Wh(\widehat\L_0)$ and 
$\log\L_0^2=\half\log\widehat\L_0^2$, as well as $\sum_i S_i=t$, yields
\be\label{Fderrel}
{\del\cF_0\over \del S_1} = {\del\cF_0\over \del S_2}
=\half {\del\cFh_0\over \del t} \quad {\rm at}\quad 
S_1=S_2={t\over 2},\ \ S_0=0 \ .
\ee

Finally, we need the integrals of $W(x) y(x)\d x$ over the $A_i$ cycles. By
(\ref{wqgrel}) and (\ref{yyhat}) they are immediately given by
\be\label{wyintwhyhint}
\int_{A_i} W(x) y(x) \d x = {1\over 4} \int_A \Wh(\xi) \yh(\xi) \d\xi \ ,
\quad i=1,2\ .
\ee
If we combine (\ref{sqsg}), (\ref{Fderrel}) and (\ref{wyintwhyhint}) and use
(\ref{homrelbis}), we conclude that  
the planar free energy $\cF_0$ of the matrix model with the even 
quartic potential (\ref{quartic}) and the planar free energy $\cFh_0$ 
of the gaussian matrix model (\ref{gaussianpot}) are related as
\ba\label{f0fohatqg}
\cF_0(S_0,S_1,S_2)\Big\vert_{S_0=0, S_1={t\over 2},  S_2={t\over 2}}
&=& \half \cFh_0(t) 
\nonumber\\
&=&{1\over 2} \left[
{t^2\over 2} \log t -{3\over 4} t^2 - t\, \Big(2b-{a^2\over 2}\Big) \right] \ ,
\ea
where we used (\ref{n=1int}) and $w_0=2b-{a^2\over 2}$.

Although (\ref{f0fohatqg}) is a nice result, it only gives the prepotential
$\cF_0$ on the submanifold of the moduli space where 
$S_1=S_2={t\over 2},\ S_0=0$. 
However, we now turn to the computation of the effective 
superpotential (\ref{weff}) and we will show that we can find vacuum
configurations on this special submanifold. They are given by the
points where  $t$ takes 
one of its vacuum values as determined by the Veneziano-Yankielowicz 
superpotential
\be\label{VY}
\Wheff(N,t)=-N{\del\cFh_0\over \del t}(t) 
+ \widehat\alpha(\widehat\L,N)\, t \ .
\ee
The vacua $\langle S_i\rangle$ are determined 
as extrema of $\Weff$, i.e. 
\be\label{vaccond}
{\del\over \del S_i} \Weff (N_i,S_i) \Big\vert_{S_i=\langle S_i\rangle}
=0 \quad \forall\ i=1,\ldots n \ .
\ee
As to find these vacua one must compute all derivatives, 
one would therefore expect that the knowledge of $\cF_0$ or $\Weff$ on a
particular submanifold of the moduli space is not enough.
We will now show that one can nevertheless find certain vacua.
To do so, it will be important to show that the $\Z_2$ symmetry is 
realised in the effective theory as the symmetry under the exchange 
of $S_1$ and $S_2$.

First of all, note that $\Weff$ and the vacua depend on the $N_i$ which
can be interpreted as the numbers of $D5$-branes wrapping the 
$i^{\rm th}$ two-cycle before the geometric transition. 
On the other hand, $S_i=t \bar N_i$ where $\bar N_i$ counts the number 
of topological branes on the $i^{\rm th}$ two-cycle. There is no 
direct relation between the $N_i$ and the $\bar N_i$ or $S_i$, except when 
some $N_i=0$. Then there are no $D5$-branes and hence no corresponding 
topological branes and no 
corresponding gauge group $U(N_i)$, and the associated $S_i$ must 
vanish. Thus if we choose $N_0=0$ then $S_0=0$ is fixed and cannot be varied.

Furthermore, we assume that $N$ is even and make the symmetric choice
$N_1=N_2={N\over 2}$. Then, according to 
the remarks below eq. (\ref{weff}), we have also 
$\a_1(\L;0,{N\over 2},{N\over 2})=\a_2(\L;0,{N\over 2},{N\over 2})$
$\equiv \alpha^{(3)}(\L;{N\over 2})$ and
\be\label{quarticWeff}
\Weff\left(0, {N\over 2},{N\over 2};0,S_1,S_2\right)
=- {N\over 2} \left[ {\del \cF_0\over \del S_1} 
+ {\del \cF_0\over \del S_2} \right] (0,S_1,S_2) 
+ \alpha^{(3)}\Big(\L;{N\over 2}\Big) \left[ S_1+S_2\right] \ .
\ee
In the following, we are interested in the dependence of this 
function on $S_1$ and $S_2$ and we often simply write $\Weff(S_1,S_2)$.
The $\Z_2$ symmetry of the original $U(N)$ gauge symmetry acts on 
$\Phi$ as $\Phi\to-\Phi$ and, in particular, it must permute the 
eigenvalues of $\Phi$ sitting close to the two solutions at $\pm\sqrt{a}$ of 
$W'(\Phi)=0$. Hence it is natural to expect that 
in the effective theory this $\Z_2$ symmetry exchanges 
$S_1$ and $S_2$. We now show that $\Weff$ indeed is 
symmetric under interchange of $S_1$ and $S_2$. This is obviously true 
for the second term and has only to be shown for the first one. To get 
$S_2\ne S_1$, but keep $S_0=0$, we must start with a more general 
$f(x)=-4tx^2+f_1 x+f_0$, restricted in such a way that $y^2(x)$ still has 
a double zero,\footnote{
For small $S_1-S_2$ we also have small $f_1$ and $f_0$ and the double 
zero of $y^2$ only moves slightly away from $x=0$, so that, to first order, 
the condition for the double zero is $f_1^2 \simeq 4(a^2-4t)f_0$ .
}
although the latter will no longer be at $x=0$.
The general picture is still given by the left part of Fig.\ 2 but now the 
two cuts have different lengths and orientations. Consider also a second 
Riemann surface $\wt\cR$ given by a $\yt(x)$ with the same $W'(x)$ but with 
$\wt f(x)=-4t x^2-f_1 x + f_0$ (i.e. $\wt t=t$, $\wt f_1=-f_1$, 
$\wt f_0=f_0$). Then, obviously, $\yt^2(x)=y^2(-x)$ 
and, actually, $\yt(x)=-y(-x)$ so that
\be\label{yytrel}
y(x')\d x'=\yt(x)\d x \ , \quad x'=-x \ .
\ee
Of course, the cuts of $\yt$ are not the same as those of $y$ 
(actually they got exchanged), 
but we continue to call ${\cal C}_1$ the cut associated with the 
critical point $x=\sqrt{a}$ 
with corresponding cycles $A_1$ and $B_1$, and to call ${\cal C}_2$ the 
cut associated with the 
critical point $x=-\sqrt{a}$ with corresponding cycles $A_2$ and $B_2$.
So we keep the same basis of $A_i$ and $B_i$ cycles for both manifolds,
according to Fig.\ \ref{quarticfig}. The map $x\to x'=-x$ then 
exchanges $A_1$ and $A_2$ as well as $B_1$ and $\tilde B_2$.
It follows for the integrals over the $A_i$ cycles that
\be\label{sstrel}
\St_1={1\over 4\pi i}\int_{A_1} \yt(x)\d x
={1\over 4\pi i}\int_{A_2} y(x')\d x' = S_2
 \qquad {\rm and} \qquad \St_2= S_1 \ .
\ee
Since the coefficients $t$, $f_1$ and $f_0$ are determined by 
$S_1,\ S_2$ (and $S_0=0$) we write
\be\label{yyts}
y(x)\equiv y(x;S_1,S_2) \ , \quad \yt(x)\equiv y(x;S_2,S_1) \ ,
\ee
($S_0=0$ is understood throughout), and (\ref{yytrel}) then reads
\be\label{yxmxs}
y(x';S_1,S_2)\d x' = y(x;S_2,S_1)\d x \ ,  \quad x'=-x \ .
\ee
It then follows, with $\wt B_2$ as in Fig.\ \ref{B1cycle},
\be\label{bys}
\int_{\wt B_2} y(x;S_2,S_1)\d x = \int_{B_1} y(x';S_1,S_2)\d x' \ .
\ee
Now it is still true that the integral over the $\wt B_2$ cycle equals 
the one over the $B_2$ cycle (since the $C_\pm$ integrals still cancel each 
other), and hence by (\ref{Biint}) we have
\be\label{dfss}
{\del \over \del s_2} \cF_0(s_1,s_2)\Big\vert_{s_1=S_2,\ s_2=S_1}
={\del \over \del s_1} \cF_0(s_1,s_2)\Big\vert_{s_1=S_1,\ s_2=S_2} \ .
\ee
Similarly,\footnote{
In the following we adopt the convention that ${\del \cF_0\over \del S_1}$
always means the derivative of $\cF_0$ with respect to its 
{\it first argument}, etc, so that eq. (\ref{dfss}) can be simply written as
${\del \cF_0\over \del S_2}(S_2,S_1) = {\del \cF_0\over \del S_1}(S_1,S_2)$.
}
one has 
${\del \cF_0\over \del S_1}(S_2,S_1) = {\del \cF_0\over \del S_2}(S_1,S_2)$,
and it follows that 
$\left({\del \cF_0\over \del S_1} 
+ {\del \cF_0\over \del S_2}\right)(S_1,S_2)$ 
is symmetric under interchange of $S_1$ and $S_2$. Hence we have shown 
that $\Weff$ of (\ref{quarticWeff}) is a symmetric function of $S_1$ and $S_2$.
Note that this is only true because $W(x)$ is an even function of $x$.

Now, as for any {\it symmetric} function of two variables, 
${\del\over \del S}\Weff(S,S)\vert_{S=S^*}=0$ implies the vanishing 
of {\it both} partial derivatives at the symmetric point:\footnote{
The proof is easy: suppose $g(x,y)=g(y,x)$ and 
$g(x,x)$ finite. We introduce $u=x+y$ and $v=x-y$. 
Then $g$ is even under $v\to -v$, and $\del_v g$ is necessarily odd and 
hence vanishes at $v=0$: $\del_v g\vert_{x=y}=0$. Also 
$2\del_u g\vert_{x=y} 
= \left[\del_x g(x,y)+\del_y g(x,y)\right]\vert_{x=y}
={\d\over \d x} g(x,x)$. Hence, ${\d\over \d x} g(x,x)\vert_{x=x^*} =0$
implies $\del_u g=\del_v g=0$ at $x=y=x^*$ and hence $\del_x g=\del_y g=0$
at $x=y=x^*$.
}
\be\label{dswdsw}
{\del\over \del S}\Weff(S,S)\Big\vert_{S=S^*}=0
\quad \Rightarrow \quad
{\del\over \del S_1}\Weff(S_1,S_2)\Big\vert_{S_1=S_2=S^*}=
{\del\over \del S_2}\Weff(S_1,S_2)\Big\vert_{S_1=S_2=S^*}=0 \ .
\ee
Thus, to find vacua of the gauge theory, a sufficient condition 
is extremality of
\be\label{weffgauss}
\Weff\left(0,{N\over 2},{N\over 2};0,{t\over 2},{t\over 2}\right) 
= - {N\over 2} {\del \cFh_0\over \del t}(t) 
+ \alpha^{(3)}\Big(\L;{N\over 2}\Big)\, t 
= {1\over 2} \Wheff(N,t)
\ ,
\ee
where we used the relations (\ref{sqsg}) and (\ref{Fderrel}), and also
identified
$\alpha^{(3)}(\L;{N\over 2}) =\half\, \widehat\alpha(\widehat\L,N)$.
Note that
${\d\over \d t}\Wheff(N,t)\Big\vert_{t=t^*}=0$ precisely gives
the Veneziano-Yankielowicz vacua. We conclude that
\be\label{dswgdsw}
{\d\over \d t}\Wheff(t)\Big\vert_{t=t^*}=0
\quad \Rightarrow \quad
{\del\over \del S_1}\Weff(S_1,S_2)\Big\vert_{S_1=S_2=t^*/2}=
{\del\over \del S_2}\Weff(S_1,S_2)\Big\vert_{S_1=S_2=t^*/2}=0 \ .
\ee
Thus, we get vacuum configurations for the $U(N/2)\times U(N/2)$ gauge 
theory with a quartic tree-level superpotential from each of the 
Veneziano-Yankielowicz vacua.
Of course, this only gives the ``symmetric'' vacua. We will have nothing 
to say for breaking patterns with $N_1\ne N_2$ or $N_0\ne 0$. 
Moreover, even for $N_1=N_2=N/2,\ N_0=0$, we expect other vacua 
also at $S_1 \ne S_2$.

Finally note that not only the effective superpotential (for 
$N_1=N_2, N_0=0$) is a symmetric function of $S_1$ and $S_2$, but 
the prepotential itself has the same symmetry (for $S_0=0$):
\be\label{f0sym}
\cF_0(0,S_2,S_1) = \cF_0(0,S_1,S_2) \ .
\ee
To see this, one uses (\ref{yxmxs}) again to show that
\be\label{ywrel}
\int_{A_2} y(x;S_2,S_1) W(x)\d x=\int_{A_1} y(x;S_1,S_2) W(x)\d x
\ee
and hence, together with (\ref{dfss}), eq. 
(\ref{homrelbis}) yields (\ref{f0sym}).
\vskip 4.mm

\subsection{General even superpotential}

%
\begin{figure}[h]
\centering
\includegraphics[width=1.0\textwidth]{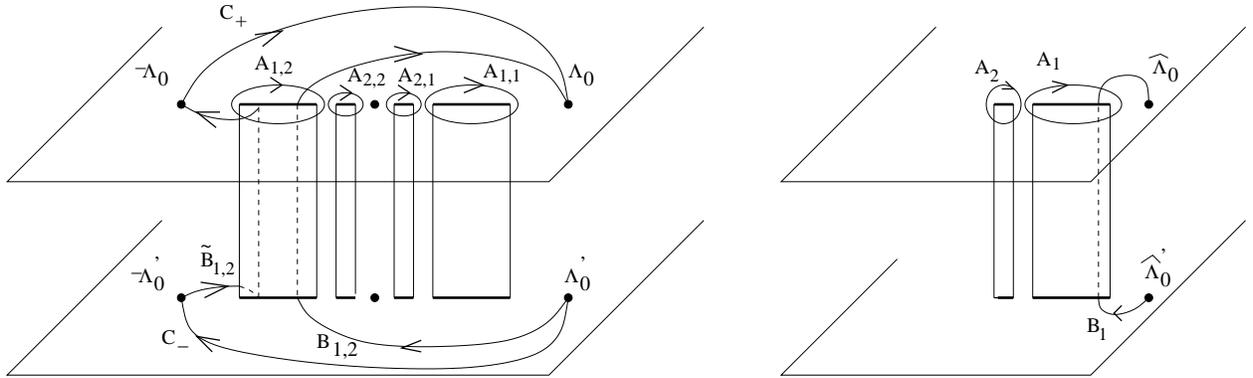}\\
\caption[]{On the left, we have depicted the Riemann surface $\cR$ for 
an even superpotential of degree 6 ($m=2$), together with some cycles. 
On the right, we show the Riemann surface $\cRh$ for the corresponding cubic 
superpotential.} \label{evenmtwo}
\end{figure}
\begin{figure}[h]
\centering
\includegraphics[width=1.0\textwidth]{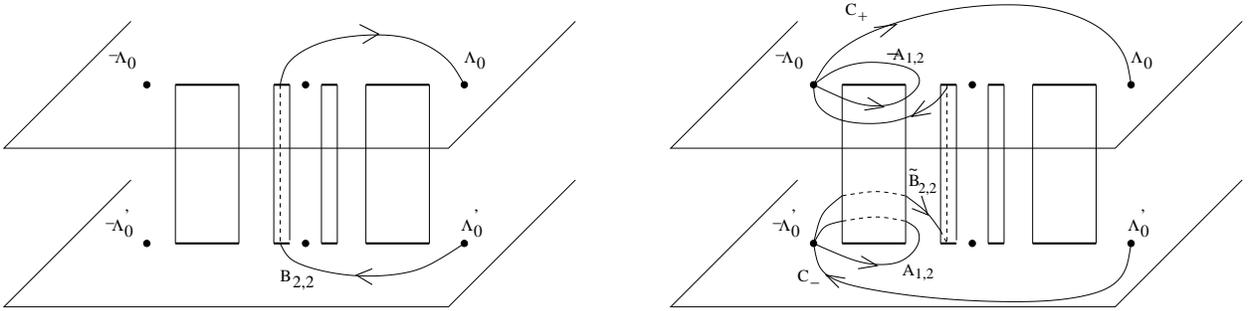}\\
\caption[]{The decomposition of the $B_{2,2}$ cycle into $C_\pm$, 
$\pm A_{1,2}$ and $\tilde B_{2,2}$ is shown.} 
\label{evenmtwoBcycle}
\end{figure}

\vskip-3.mm
Next, we consider the case of a general even superpotential 
$W(x)$ of order $2m+2$. Being even, we can always write
\be\label{wwhgen}
W(x)=\half \Wh(\xi) \ , \quad \xi=x^2 \ ,
\ee
where $\Wh(\xi)$ now is of order $m+1$. If furthermore we choose 
$f(x)=x^2 \widehat f(\xi)$ we have for $y^2(x)=W'(x)^2+f(x)$ and 
$\yh^2(\xi)=\Wh'(\xi)^2+\widehat f(\xi)$ the same relation as before, 
namely $y(x)\d x =\half \yh(\xi)\d\xi$. Again, we call $\cR$ and $\cRh$ 
the Riemann surfaces corresponding to $y$ and $\yh$, respectively.
We label the cuts and cycles in such a way that the $A_{l,1}$ and 
$A_{l,2}$ cycles of $\cR$ are mapped 
to the $A_l$ cycle of $\cRh$ for all 
$l=1,\ldots,m$, see Fig.\ \ref{evenmtwo}. The cut that does not open 
is again labelled by $0$.
Hence
\be\label{specialp}
S_{l,1}=S_{l,2}=\half \Sh_l \ , 
\quad l=1,\ldots m \ , \quad S_0=0 \ .
\ee
In particular, $t=\widehat t$. For the $B$ cycles, the 
$B_{l,1}$ cycles of $\cR$ are  directly mapped to the 
$B_l$ cycles of $\cRh$, while for the $B_{l,2}$ cycles things are
more subtle. They have to be decomposed into  cycles 
$\tilde B_{l,2}$ which are mapped to $B_l$, as well as various other pieces.
For example,  $B_{1,2}$ is decomposed as 
$B_{1,2} = C_- + \wt B_{1,2} + C_+$ as shown 
in Fig.\ \ref{evenmtwo}, with 
the integrals over $C_-$ and $C_+$ cancelling each other and
$\wt B_{1,2}$ being mapped to $B_1$. The $B_{2,2}$ cycle is decomposed
as follows
(see Fig.\ \ref{evenmtwoBcycle}).
\vskip-3.mm
\noindent
One first goes on a large arc $C_-$ on the lower sheet to $-\L_0'$ and 
from there one must encircle the cut ${\cal C}_{1,2}$ counterclockwise 
(which is homologous to $A_{1,2}$) before going on $\tilde B_{2,2}$ 
through the cut ${\cal C}_{2,2}$ to $-\L_0$ on the upper sheet. There 
again one has to encircle the cut ${\cal C}_{1,2}$ counterclockwise 
(which on the upper sheet is homologous to $-A_{1,2}$), before going 
on the large arc $C_+$ to $\L_0$. A similar decomposition applies 
for all $B_{l,2}$:
\be\label{bbtildedecompeven}
B_{l,2}=C_- + \sum_{l'=1}^{l-1}  A_{l',2}
+\wt B_{l,2} - \sum_{l'=1}^{l-1}  A_{l',2} + C_+ \ .
\ee
Of course, the integrals over the $A_{l',2}$ cancel, as do those over 
$C_\pm$, while $\tilde B_{l,2}$ is mapped to $B_l$.
Thus the $B_{l,2}$ integrals equal the $\tilde B_{l,2}$ integrals for 
all $l$ and $\int_{B_{l,p}} y(x)\d x={1\over 2} \int_{B_l} \yh(\xi)\d\xi$.

As a result, 
one concludes, as for the quartic superpotential, that
\be\label{f0dergen}
{\del \cF_0\over \del S_{l,1}} = {\del \cF_0\over \del S_{l,2}}
= \half {\del \cFh_0\over \del \Sh_l} \quad {\rm at} \quad
S_{l,1}=S_{l,2}=\half \Sh_l \ , \ \ S_0=0  \ .
\ee
Also $\int_{A_{l,1}} W(x) y(x)\d x = \int_{A_{l,2}} W(x) y(x)\d x 
={1\over 4} \int_{A_l} \Wh(x) \yh(\xi)\d\xi$, and by (\ref{homrelbis}) 
one has 
\be\label{ffhgen2}
\cF_0(0,S_{l,1},S_{l,1})
=\half \cFh_0(\Sh_l) \ , 
\quad S_{l,1}=\half \Sh_l \ .
\ee
In general, however, we do not have explicit expressions\footnote{
As noted in the introduction, for $m=2$, one can still, in principle, 
express $\cF_0$ through various combinations of incomplete elliptic 
functions
} 
for $\cFh_0$, contrary to the case $m=1$.

We can exploit further the fact that $W(x)$ is an even function of 
$x$ to show relations analogous to (\ref{yxmxs}), (\ref{bys}) and 
(\ref{ywrel}). For these relations to be true it is crucial that 
$\int_{\tilde B_{l,2}} y(x)\d x = \int_{B_{l,2}} y(x)\d x$ even 
for $S_{l,1}\ne S_{l,2}$. From our discussion above this is obviously 
the case. We conclude that
\be\label{fsymgen2}
\cF_0(0,S_{l,1},S_{l,2})
=\cF_0(0,S_{l,2},S_{l,1}) \ .
\ee

We can also compute the effective superpotential $\Weff$ on the
submanifold (\ref{specialp}) of the moduli space and relate it to $\Wheff$:
\be\label{W-Wh}
\Weff\left(0,{N_l\over 2},{N_l\over 2};0,{\Sh_l\over 2},{\Sh_l\over 2}\right)
=\half \Wheff(N_l,\Sh_l) \ .
\ee
However, we are not able to show that the vacua of $\Wheff$ correspond
to (some of the) vacua of $\Weff$, although this might be expected to be true. 
Indeed, to prove this would require to show that  the $2m$ derivatives 
of $\Weff$ vanish, while extremality of $\Wheff$ only gives $m$ 
conditions, and the symmetry of $W$ only forces one more derivative 
to vanish. It is only for $m=1$ that we have the right number of conditions.

\subsection{Superpotentials of degree $2k$ with $\Z_k$-symmetry}

%
\begin{figure}[h]
\centering
\includegraphics[width=0.5\textwidth]{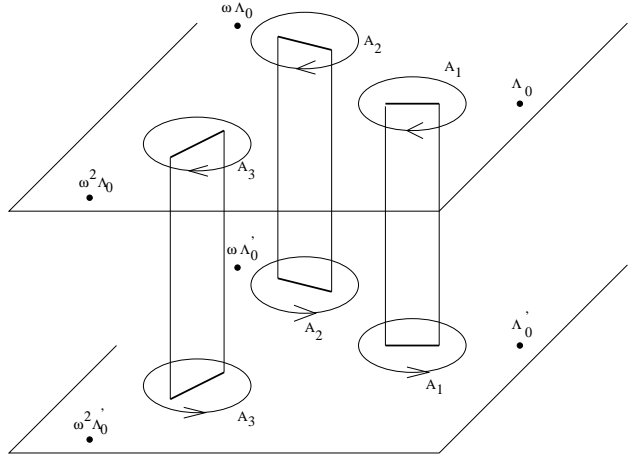}\\
\caption[]{
Shown is the Riemann surface $\cR$ for $m=1$, $k=3$ with 
its three non-degenerate cuts ${\cal C}_1,\ {\cal C}_2$ and ${\cal C}_3$ 
connecting the two sheets and the 
$A_i$ cycles surrounding these cuts. We do not show the degenerate cuts 
corresponding to the double zeros of $y^2$.
Note again that a clockwise oriented 
cycle on the upper plane is homologous to a counterclockwise oriented 
cycle on the lower plane. We also show the marked 
points $\L_0$ and $\L_0'$ on the upper and lower sheet as well as
these points rotated by $\o=e^{2\pi i/3}$ and by $\o^2=e^{-2\pi i/3}$.
} \label{x3}
\vskip-3.mm
\end{figure}

Now we want to consider the general cases with $\Z_k$ symmetry.
Start with a $W(x)$ of order $2k$, $k\ge 3$, 
having a $\Z_k$-symmetry generated 
by $x\to \o x$, $\o=e^{2\pi i/k}$. This is necessarily of the form
\be\label{wk}
W(x)={1\over 2k}x^{2k} -{a\over k} x^k + b \ .
\ee
We let
\be\label{xixk}
\xi=x^k \ , \quad W(x)={1\over k} \Wh(\xi) \ , 
\quad \Wh(\xi)=\half(\xi-a)^2 + w_0 \ ,
\ee
where $w_0=k b -{a^2\over 2}$. Much of the discussion is analogous to the 
case of the quartic superpotential, but there are also some
important differences that appear for $k\ge 3$.

We have $W'(x)=x^{k-1}(x^k-a)$ and choose $f(x)=-4t x^{2k-2}$, i.e. 
$f_0=f_1=\ldots=f_{2k-1}=0$. Then 
$y^2(x)=W'(x)^2+f(x) = x^{2(k-1)}\left[ (x^k-a)^2-4t\right]
\equiv x^{2(k-1)} \yh^2(x^k)$ and one gets $k$ cuts 
${\cal C}_1,\ldots {\cal C}_k$, as well as 
a multiple zero at $x=0$. The latter corresponds to $k-1$ degenerate 
cuts ${\cal C}_{0,1},\ldots, {\cal C}_{0,k-1}$, on top of each other. 
The non-degenerate cuts and the corresponding $A_1,\ldots A_k$ cycles 
are shown in Fig.\ \ref{x3} for $k=3$ (and $0<2\sqrt{t}<a$). 
All these $A_1,\ldots A_k$ cycles are 
mapped onto the single $A$ cycle in the $\xi$-plane. We have
\be\label{yyhk}
y(x)\d x={1\over k}\yh(\xi)\d\xi \ ,
\ee
and
\be\label{sik}
\int_{A_i}y(x)\d x={1\over k} \int_A \yh(\xi) \d\xi \quad
\Rightarrow\quad S_i={t\over k} \ , \quad i=1,\ldots k \ .
\ee
In particular, we have again $t=\sum_{i=1}^k S_i = \wh t$. There are 
also $k-1$ vanishing $S$'s corresponding to the multiple zero at $x=0$. 
We will 
denote them
\be\label{si0}
S_{0,1}=\ldots =S_{0,k-1}=0 \ .
\ee

\begin{figure}[h]
\centering
\includegraphics[width=1.0\textwidth]{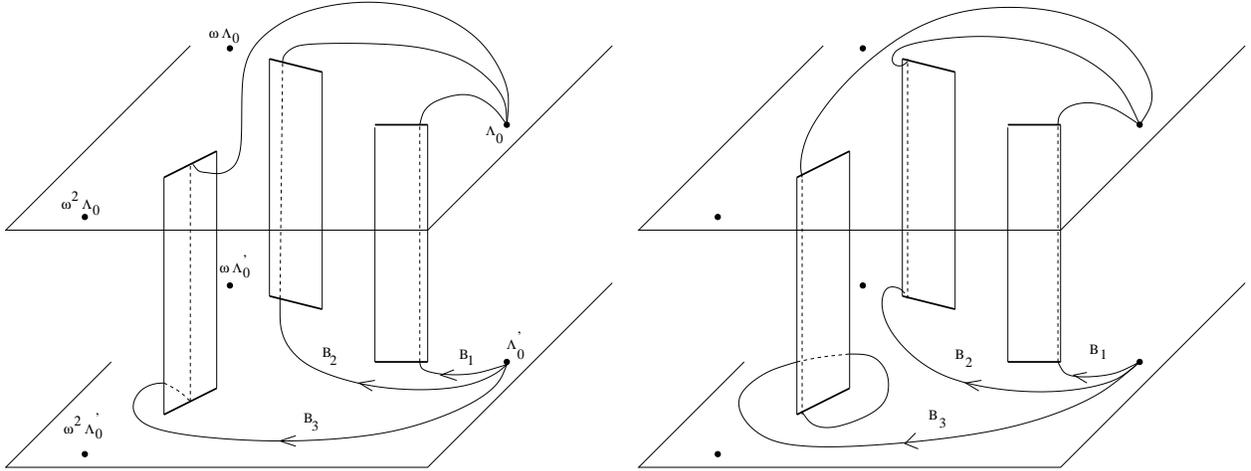}\\
\caption[]{
In the left figure we display a certain choice of 
$B_i$ cycles that begin at $\L_0'$ on the lower sheet, go through 
the  cut ${\cal C}_i$ and end at $\L_0$ on the upper sheet. The right
figure shows a choice of $B_i$ cycles 
homologous to the one of the left figure. To see this 
one has to 
remember that a given side of a cut on the lower sheet is identified 
with the opposite side of the cut on the upper sheet.
} \label{x3-12}
\vskip-3.mm
\end{figure}

\begin{figure}[h]
\centering
\includegraphics[width=1.0\textwidth]{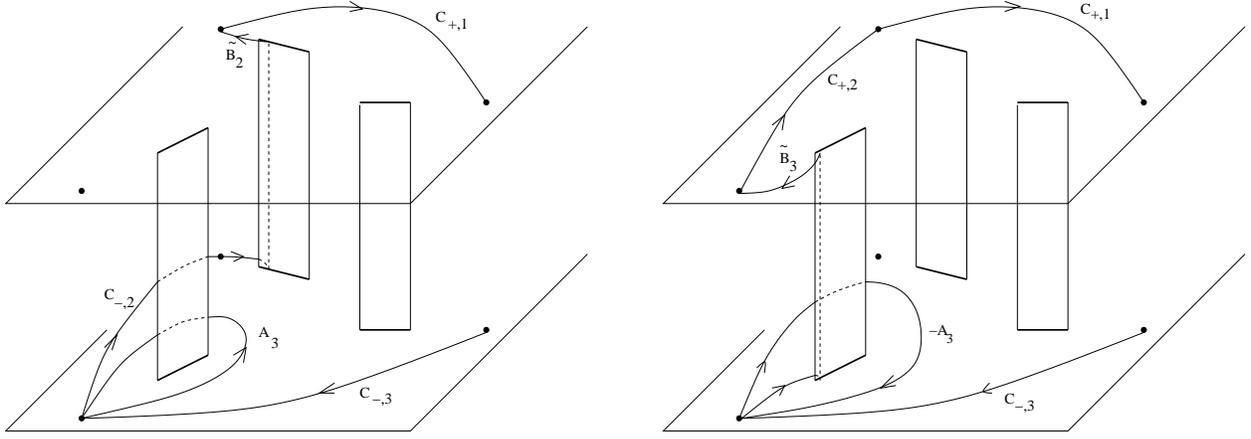}\\
\caption[]{
The left part of this figure concentrates on the $B_2$ cycle. 
It shows that it is homologous to a cycle that runs on a large arc from 
$\L_0'$ on the lower sheet to $\o^2\L_0'$, then encircles the cut 
${\cal C}_3$ counterclockwise returning 
to $\o^2\L_0'$, from there goes on a large arc 
to $\o\L_0'$,
then goes from $\o\L_0'$ through the cut ${\cal C}_2$ to 
$\o\L_0$ on the upper sheet, and from there on a large arc to $\L_0$.
The right part concentrates on the $B_3$ cycle.
and shows that it is homologous to a cycle 
that runs on a large arc from 
$\L_0'$ on the lower sheet to $\o^2\L_0'$ ($C_{-,3}$), then encircles 
the cut ${\cal C}_3$
clockwise returning to $\o^2\L_0'$, then goes from $\o^2\L_0'$ through 
this same cut ${\cal C}_3$ to $\o^2\L_0$ on the upper sheet ($\wt B_3$), 
and from there on a large arc $C_{+,2}$ to $\o\L_0$, and then on $C_{+,1}$
to $\L_0$.} \label{x3-45}
\vskip-3.mm
\end{figure}

The integrals over the $B_i$ cycles involve some subtleties, not present 
for $k=2$. To see this, concentrate first on $k=3$ and consider the choice of 
$B_i$ cycles shown in Fig.\ \ref{x3-12} (consistent with Fig.\ \ref{x3-0}).
We want to see whether or not these cycles are mapped to the $B$ cycle 
on the Riemann surface $\cRh$. This is obvious for $B_1$, 
but less obvious for $B_2$ and $B_3$, which must first be
decomposed into various pieces.
As shown in the left part of Fig.\ \ref{x3-45}, 
the decomposition of $B_2$ is
\be\label{B2}
B_2 \simeq C_{-,3}+ A_3 + C_{-,2}  + \wt B_2 + C_{+,1} \ ,
\ee
where $C_{-,3}$ is
a large arc going from 
$\L_0'$ to $\o^2 \L_0'$ on the lower sheet, $C_{-,2}$ another large arc 
going from $\o^2\L_0'$ to $\o \L_0'$ still on the lower sheet; 
$\wt B_2$ goes from $\o \L_0'$ through the cut ${\cal C}_2$ to 
$\o\L_0$ on the upper sheet, and $C_{+,1}$ is a large arc  on 
the upper sheet from $\o\L_0$ to $\L_0$. 
Now $C_{-,3}$ and $C_{-,2}$ are both mapped to $-A$, while $A_3$ and 
$C_{+,1}$ are both mapped to $A$, and $\wt B_2$ is mapped to $B$. 
Hence the integrals over $C_{-,3}$,
$C_{-,2}$, $A_3$ and $C_{+,1}$ cancel and
\be\label{B2int}
\int_{B_2} y(x)\d x = \int_{\wt B_2} y(x)\d x 
={1\over 3} \int_B \yh(\xi) \d\xi \ .
\ee
Similarly, if we choose the $B_3$ cycle as shown in 
Fig.\ \ref{x3-12}
it can be decomposed as in the right part of Fig.\ \ref{x3-45}:
\be\label{B1}
B_3=C_{-,3} -A_3+\wt B_3 +C_{+,2}+C_{+,1} \ ,
\ee
where $\wt B_3$ goes from $\o^2\L_0'$ through the cut ${\cal C}_3$ to 
$\o^2\L_0$ and $C_{+,2}$ from $\o^2\L_0$ to $\o\L_0$. Again, $\wt B_3$
is mapped to $B$, while $C_{-,3}$ 
and $-A_3$ are mapped to $-A$, and $C_{+,2}$ and $C_{+,1}$ are mapped 
to $A$, so that the corresponding integrals cancel. The result is
\be\label{B1int}
\int_{B_3} y(x)\d x = \int_{\wt B_3} y(x)\d x 
={1\over 3} \int_B \yh(\xi) \d\xi \ .
\ee
Note that one could have made a  {\it choice} for $B_3$ different from 
the one shown in Fig.\ \ref{x3-12}, e.g. not to 
first encircle the cut ${\cal C}_3$ on the lower sheet.
Then one would have missed the $-A_3$ piece, resulting in an additional 
piece $\int_{A_3} y(x)\d x$ on the r.h.s of (\ref{B1int}). As discussed in
section 2, such a 
different choice is always possible as it corresponds to the symplectic 
change of basis $B_3\to B_3 + A_3$ and results in an additional piece
$i \pi  S_3^2$ in the prepotential.  However, this extra piece spoils 
the ``reality'' of $\cF_0$ and, more importantly, it would spoil the 
symmetry of $\cF_0$ under exchange of the $S_1$, $S_2$ and $S_3$ to 
be discussed below. We conclude, that it is important for us to
make precisely the choice of $B_i$ cycles shown in Fig.\ \ref{x3-12}.

\begin{figure}[h]
\centering
\includegraphics[width=0.7\textwidth]{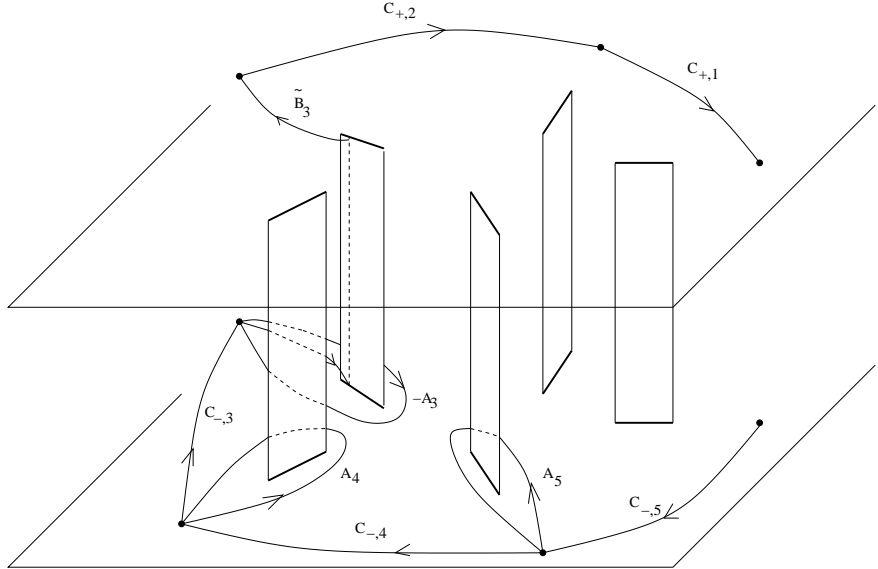}\\
\caption[]{
This figure shows, for $k=5$ and $i=3$ how our choice of
$B_i$ cycles is decomposed into large arcs $C_{\pm,p}$ running from 
$\o^{p}\L_0$ to $\o^{p-1} \L_0$ on the lower or upper sheet, various $A_q$
cycles and the $\tilde B_i$ cycle.
} \label{x3-9}
\end{figure}

In fact, this is easily generalised to arbitrary $k$. We can always 
consistently 
deform our $B_i$ cycles into a sum of large arcs $C_{\pm,p}$ running from 
$\o^{p}\L_0$ to $\o^{p-1} \L_0$ on the lower or upper sheet, various $A_j$
cycles and a $\tilde B_i$ cycle, see Fig.\ \ref{x3-9}. 
More precisely, on the lower sheet we 
start at $\L_0'$ and run on a large arc $C_{-,k}$ to $\o^{k-1}\L_0'$. Then 
we encircle the cut ${\cal C}_k$ counterclockwise, which is homologous to 
$A_k$. Next, we go on another arc $C_{-,k-1}$ from 
$\o^{k-1}\L_0'$ to $\o^{k-2}\L_0'$, encircle 
the cut ${\cal C}_{k-1}$ counterclockwise, and so on, until we reach 
$\o^{i-1}\L_0'$ which is the starting point of $\tilde B_i$.
So far there was no arbitrariness. Now we first encircle the cut 
${\cal C}_i$ {\it clockwise} $m_i$ times. This number $m_i$ is arbitrary, 
a priori, but if we fix it as $m_i=i-2$ we will obtain equality of the 
$B_i$ and $\tilde B_i$ integrals. Next, the $\tilde B_i$ cycle goes 
from $\o^{i-1}\L_0'$ through the cut
${\cal C}_i$ to $\o^{i-1}\L_0$ on the upper sheet. From there we go 
on $i-1$ large arcs $C_{+,r}$ ($r=i-1,\ldots 1$)  through 
$\o^{i-2}\L_0$, etc to $\L_0$. 
The result is, for $i=2,\ldots k$,
\ba\label{Bikdecomp}
B_i&=& (C_{-,k} +A_k) + (C_{-,k-1} +A_{k-1}) + \ldots +(C_{-,i+1} +A_{i+1})
\nonumber\\
&&+(C_{-,i} -(i-2)A_i) + \tilde B_i + \sum_{r=1}^{i-1} C_{+,r}  \ .
\ea
Each $C_{\pm,r}$ is mapped to $\pm A$, so that 
$\int_{C_{\pm,r}}y(x)\d x=\pm \int_{A_r}y(x)\d x 
= \pm {1\over k}\int_A \yh(\xi)\d\xi =\pm {4\pi i \over k} t$
and it is immediately clear that
\be\label{Bikint}
\int_{B_i} y(x) \d x = \int_{\wt B_i} y(x) \d x 
= {1\over k}\int_{B} \yh (\xi) \d \xi \ ,
\ee
where the $B$ cycle runs, of course, from $\Lh'_0=(\L'_0)^k$ to 
$\Lh_0=(\L_0)^k$. 

Using (\ref{Bikint}), as well as $W(\L_0)={1\over k} \Wh(\Lh_0)$ 
and $\log \L_0^2 = {1\over k}\log \Lh_0^2$, in eq. (\ref{Biint}) then yields
\be\label{dfdsk}
{\del \cF_0\over \del S_1 } =
{\del \cF_0\over \del S_2 } = \ldots =
{\del \cF_0\over \del S_k } =
{1\over k} {\del \cFh_0\over \del t } \quad {\rm at} \
S_1=\ldots=S_k={t\over k}\ , \quad S_{0,r}=0\ ,\ r=1,\ldots,k-1 \ .
\ee
Using (\ref{homrelbis}) with (\ref{dfdsk}), (\ref{sik}) and an 
analogous relation for the integrals of $W(x) y(x)\d x$, we finally arrive at
\ba\label{f0k}
\cF_0(S_{0,r},S_i)\Big\vert_{S_{0,r}=0,\ S_i=t/k} &=&
{1\over k} \cFh(t) \nonumber\\
&=& {1\over k} \left[
{t^2\over 2} \log t -{3\over 4} t - t \left(k b -{a^2\over 2} \right) 
\right] \ .
\ea

Our next task is to study whether for {\it different} $S_i$ the prepotential 
$\cF_0$ or the effective superpotential $\Weff$ are symmetric under cyclic 
permutations of the $S_i$. The answer will be positive for $\Weff$
allowing us to find vacua from those of the Veneziano-Yankielowicz 
superpotential. However, it will be negative for $\cF_0$ due to subtleties
in the precise definition of the $B_i$ cycles having to do with the necessity 
to choose a common cut-off $\L_0$ for all cycles $B_i$. In a sense, this 
is like an anomaly.

We will 
proceed similarly to the discussion for the even quartic superpotential. 
Now the superpotential has a $\Z_k$-symmetry $W(\o x)=W(x)$,  where
$\o=e^{2i\pi/k}$. We will compare different Riemann surfaces related 
to each other essentially by a $\Z_k$ rotation $x\to \o x$. 
The subtlety, however, is that $\L_0$ is kept fixed, and the previously 
shown (non)equivalence of the $\wt B_i$ and $B_i$ cycles will be crucial.

We start with $y^2=W'(x)^2+f(x)$ and
\vskip-3.mm
\be\label{fk}
f(x)=-4t x^{2k-2} + \sum_{p=0}^{2k-3} f_p x^p \ ,
\ee
where now the $f_p$ are non-zero but still such that $y^2$ has $k-1$ 
double zeros (although they will no longer all be at $x=0$). In 
particular, we still have
$S_{0,r}=0 \ , \ r=1,\ldots, k-1$. 
We let $\wt f_p=\o^{p+2} f_p$ so that $\wt t = t$ and, with obvious notation,
\be\label{ftilde}
\wt f(x) = \o^2 f(\o x) \ , \quad \yt^2=W'(x)^2 + \wt f(x) \ .
\ee
Then we have $\yt^2(x)=\o^2 y^2(\o x)$ and
\be\label{yto}
\yt(x) \, \d x = y(x')\, \d x' \ , \quad x'=\o x \ .
\ee
The map $x\to x'=\o x$ maps the $A_i$ cycle to the $A_{i+1}$ cycle 
($A_{k+1}\equiv A_1$), cf. Fig \ref{x3}. It follows that 
\be\label{sisi+1}
\wt S_i={1\over 4\pi i} \int_{A_i} \yt(x)\d x 
={1\over 4\pi i} \int_{A_{i+1}} y(x')\d x' = S_{i+1} \ .
\ee
Similarly, under $x\to x'=\o x$, the $\wt B_i$ cycle is mapped to the 
$\wt B_{i+1}$ cycle\footnote{
It should be clear that the $\wt B_i$ cycles are defined as in 
Figs. \ref{x3-45} and \ref{x3-9}, and, as for the discussion of the 
quartic superpotential, the tildes on the $\wt B_i$ have 
nothing to do with the tildes on $\yt$ or $\wt S_i$. We apologize for 
too many tildes!
}
(with $\wt B_{k+1}\equiv \wt B_1\equiv B_1$) and
\vskip-3.mm
\be\label{BitBi+1t}
\int_{\wt B_i} \yt(x)\d x
=\int_{\wt B_{i+1}} y(x')\d x' \ .
\ee

One has to be very careful here, since we have indeed shown equality 
of the $B_i$ and $\wt B_i$ integrals, but only at the symmetric point 
where $S_1= \ldots =S_k={t\over k}$. This is no longer true once the 
$S_i$ are allowed to take different values. Then 
$\int_{A_j} y(x)\d x = 4\pi i S_j$ while 
$\int_{C_{\pm,r}} y(x)\d x= \pm {4\pi i\over k}\, t$ from the 
asymptotics of $y$. It then  follows from
(\ref{Bikdecomp}) that
\vskip-4.mm
\be\label{BiBitdiff}
\int_{B_i}y(x)\d x=\int_{\wt B_i}y(x)\d x
+ 4\pi i \left( \sum_{j=i+1}^k S_j - (i-2) S_i + (2i-2-k){t\over k}\right) \ ,
\quad i=1,\ldots k \ .
\ee
(For $i=1$ this yields correctly $\int_{B_1}y(x)\d x=\int_{\wt B_1}y(x)\d x$.)
Although the individual differences are non-vanishing for each $i\ne 1$, their
sum vanishes and thus
\vskip-3.mm
\be\label{BiBitsum}
\sum_{i=1}^k \int_{B_i}y(x)\d x=\sum_{i=1}^k \int_{\wt B_i}y(x)\d x \ .
\ee
Of course, the same relation holds with $y$ replaced by $\yt$.
Combining (\ref{BitBi+1t}) and (\ref{BiBitsum}) then shows that
\vskip-3.mm
\be\label{BiBi+1}
\sum_{i=1}^k \int_{B_i} \yt(x)\d x
=\sum_{i=1}^k\int_{\wt B_i} \yt(x)\d x
=\sum_{i=1}^k\int_{\wt B_{i+1}} y(x')\d x'
=\sum_{j=1}^k\int_{\wt B_j} y(x')\d x'
=\sum_{j=1}^k\int_{B_j} y(x')\d x' \ .
\ee
As discussed for the quartic superpotential, the coefficients $f_p$ are 
determined by the $S_i$, and $\yt(x)$ can be rewritten as 
$\yt(x)\equiv y(x;\wt S_1, \wt S_2, \ldots \wt S_{k-1}, \wt S_k) 
\equiv y(x; S_2, S_3, \ldots S_k, S_1)$ by (\ref{sisi+1}). Then 
eq. (\ref{BiBi+1}) reads
\vskip-3.mm
\be\label{Bisi+1}
\sum_{i=1}^k\int_{B_i} y(x; S_2, S_3, \ldots S_k, S_1) \d x = 
\sum_{j=1}^k\int_{B_j} y(x; S_1, S_2, \ldots S_{k-1}, S_k)  \d x \ ,
\ee
and from eq. (\ref{Biint}) immediately
\vskip-2.mm
\be\label{dfi+1}
\sum_{i=1}^k{\del \over \del s_i} \cF_0 (s_i) \Big\vert_{s_r=S_{r+1}} =
\sum_{j=1}^k{\del \over \del s_j} \cF_0 (s_j) \Big\vert_{s_r=S_r} \ .
\ee
\vskip-2.mm
\noindent
Note that, contrary to the case $k=2$, we now have
$\sum_{i=1}^k s_i {\del \over \del s_i} \cF_0 (s_i) \Big\vert_{s_r=S_{r+1}}$ 
\break
$\ne
\sum_{i=1}^k s_i{\del \over \del s_i} \cF_0 (s_i) \Big\vert_{s_r=S_r}$, 
in general, and we can no longer use (\ref{homrelbis}) to
conclude that $\cF_0$ is symmetric 
under cyclic permutations of its arguments. Again, this is due to the 
difference of the $B_i$ and $\wt B_i$ cycles, i.e. due to the 
introduction in the quantum theory of a common cutoff $\L_0$ which 
spoils the classical $\Z_k$ symmetry: we have a ``permutation anomaly''.

Nevertheless, (\ref{dfi+1}) is all we need in order to show the corresponding 
symmetry of the effective superpotential and to be able to obtain vacua.
We choose $N_{0,s}=0$ (so that the $S_{0,s}$ 
remain zero), $N_i={N\over k}$, $i=1,\ldots k$ and denote 
$\alpha_i\equiv\alpha^{(2k-1)}(\L;{N\over k})$ so that
\vskip-5.mm
\be\label{weffk}
\Weff\left(N_{0,s}=0, N_i={N\over k}; S_{0,s}=0, S_i\right)
=\sum_{i=1}^k \left[ -{N\over k} {\del \cF_0\over \del S_i}(S_{0,s}=0, S_i)
+\alpha^{(2k-1)}\Big(\L;{N\over k}\Big) S_i \right] \ .
\ee
According to (\ref{dfi+1}) this is invariant under 
cyclic permutations of the $S_i$. Again, due to this symmetry, 
$\Weff$ has a critical point with 
respect to {\it independent} variations\footnote{
Suppose that $F(s_2,\ldots s_k,s_1)=F(s_1,s_2,\ldots s_k)$. 
A pedestrian proof that ${\d\over \d s}F(s,s,\ldots s)\vert_{s=s^*}=0$ 
implies
${\del F\over \del s_i}(s_1,\ldots s_k)\vert_{s_1=\ldots s_k=s^*}=0$, 
$\forall i=1,\ldots k$, is the following:
One changes variables to $u=\sum_{i=1}^k s_i$ and 
$v_r=\sum_{i=1}^k \o^{i r} s_i$, $r=1,\ldots k-1$. Then under 
$(s_1,s_2,\ldots s_k)\to (s_2,s_3,\ldots s_1)$ one has 
$v_r\to \o^{-r} v_r$ while $u$ is invariant. Since $F$ is invariant, it can 
depend arbitrarily on $u$, but dependence on the $v_r$ can only be 
through invariant products of the $v_r$. In particular, $F$ cannot 
depend linearly on any of the $v_r$ and thus
${\del\over \del v_r}F\vert_{v_p=0}=0$. But $v_p=0\ \forall p$ is 
equivalent to $s_1=s_2=\ldots =s_k$, and we see that at the symmetric 
point all derivatives of $F$ with respect to $v_r$ automatically 
vanish. Furthermore 
${\d\over \d s} F(s,s,\ldots s)=k {\del\over \del u} F(u,v_r)\vert_{v_p=0}$. 
Hence, vanishing of ${\d\over \d s} F(s,s,\ldots s)\vert_{s=s^*}$ implies
vanishing of all partial derivatives ${\del F\over \del u}$ and 
${\del F\over \del v_r}$ and hence of all ${\del F\over \del s_i}$  
at the point $s_1=\ldots=s_k=s^*$.
}  
of {\it all} $S_i$, $i=1,\ldots k$ if (we have identified 
$\alpha^{(2k-1)}(\L;{N\over k})={1\over k} \wh\alpha(\Lh,N)$)
\be\label{wwhk}
\Wheff(N,t) 
= k \Weff\left(N_{0,r}=0, N_i={N\over k}; S_{0,r}=0, S_i={t\over k}\right)
\ee
has a critical point with respect to $t$:
\ba\label{critical}
&&\hskip-5.mm{\d\over \d t} \Wheff(N,t)\Big\vert_{t=t^*}=0 \\
\hskip-5.mm\Rightarrow&&\hskip-5.mm
{\del\over \del S_i} 
\Weff\left(N_{0,s}=0, N_i={N\over k}; S_{0,s}=0, S_i\right)
\Big\vert_{S_1=\ldots=S_k={t^*\over k}} = 0 \ ,\quad
\forall i=1,\ldots k \ .
\ea
Thus we get vacua for the $U(N)$ gauge theory, broken to 
$\prod_{i=1}^k U({N\over k})$ by a tree-level superpotential 
of order $2k$ having a $\Z_k$ symmetry, from the Veneziano-Yankielowicz vacua!

\subsection{General superpotentials  with $\Z_k$-symmetry}

\begin{figure}[h]
\centering
\includegraphics[width=0.7\textwidth]{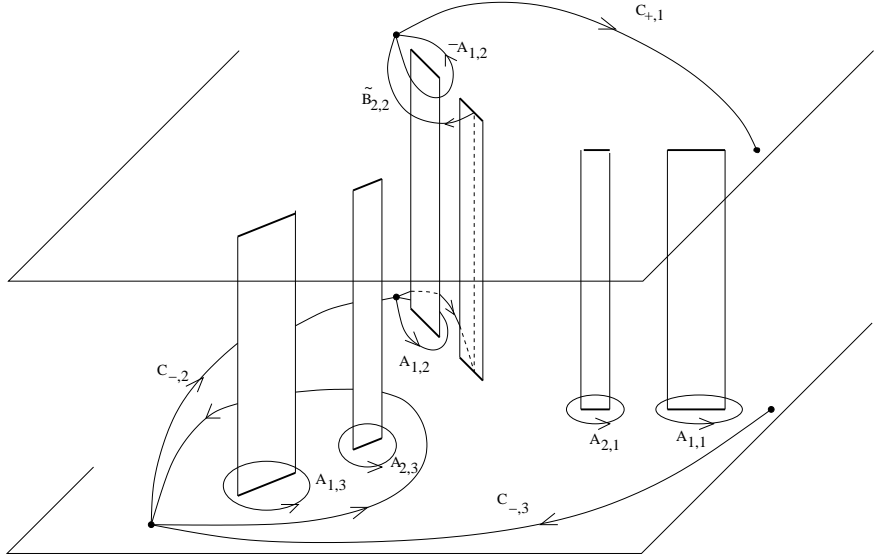}\\
\caption[]{
This figure shows, for $k=3$ and $m=2$, how the
$B_{2,2}$ cycle is  decomposed into large arcs $C_{\pm,p}$, 
various $A$ cycles and the  $\tilde B_{2,2}$ cycle. The decomposition of
$B_{1,2}$ is the same except that, once at $\o\L_0'$, one goes directly on the 
$\tilde B_{1,2}$ cycle through the cut ${\cal C}_{1,2}$ to $\o\L_0$, without 
encircling the cut on the $A_{1,2}$ cycle.
} 
\label{x3-10}
\end{figure}

A general superpotential with a $\Z_k$-symmetry is a polynomial in 
$\xi=x^k$ of order $m+1$,
\be\label{wzkgen}
W(x)={1\over (m+1)k}x^{(m+1)k} + \sum_{r=0}^m {g_{rk}\over r k} x^{rk} \ ,
\ee
and it is mapped to a corresponding ${1\over k} \Wh(\xi)$ of order $m+1$. 
If we restrict to $f(x)$ of the form $f(x)=x^{2k-2} \widehat f(\xi)$ we have
\be\label{ykgen}
y^2(x)=W'(x)^2 +f(x)=x^{2(k-1)} \left[ \Wh'(\xi)^2+\widehat f(\xi) \right]
\equiv x^{2(k-1)} \yh^2(\xi) \ .
\ee
Now, the Riemann surface $\cRh$ has $m$ cuts with 
$A$ cycles 
$A_l,\ l=1,\ldots m$ and corresponding $B$ cycles $B_l$, while the Riemann 
surface $\cR$ has $km$ (non-degenerate) cuts with $A$ cycles $A_{l,p}$ 
and $B$ cycles 
$B_{l,p}$ such that all $A_{l,p}$, $p=1,\ldots k$ are mapped to $A_l$. 
For the $B_{l,p}$ cycles one must first decompose them into various 
large arcs, $A$ cycles and a $\widetilde B_{l,p}$ cycle. This is shown 
in Fig.\ \ref{x3-10} for $k=3$ and $m=2$.
The precise choice of the $B_{l,p}$ cycles is given by a straightforward 
generalisation of eqs. (\ref{Bikdecomp}) and (\ref{bbtildedecompeven}), namely
for $p=2,\ldots k$
\ba\label{BBtildekgen}
B_{l,p}&=& \left(C_{-,k} +\sum_{q=1}^m A_{q,k}\right) 
+ \left(C_{-,k-1} +\sum_{q=1}^m A_{q,k-1}\right) + \ldots 
+\left(C_{-,p+1} +\sum_{q=1}^m A_{q,p+1}\right)
\nonumber\\
&&+\left(C_{-,p} -(p-2)\sum_{q=1}^m A_{q,p}\right) 
+\sum_{q=1}^{l-1} A_{q,p}
+ \tilde B_{l,p} 
-\sum_{q=1}^{l-1} A_{q,p}
+ \sum_{r=1}^{p-1} C_{+,r}  \ .
\ea
Then we have
\be\label{sshkgen}
S_{l,p}={1\over 4\pi i} \int_{A_{l,p}} y(x)\d x = 
{1\over 4\pi i} {1\over k}\int_{A_l} \yh(\xi)\d\xi =
{1\over k} \Sh_l \ ,
\ee
and
\be\label{BBkgen}
\int_{B_{l,p}} y(x)\d x = \int_{\tilde B_{l,p}} y(x)\d x =
{1\over k}\int_{B_l} \yh(\xi)\d\xi 
\quad \Rightarrow \quad
{\del\cF_0\over \del S_{l,p}} = {1\over k} 
{\del\cFh_0\over \del \Sh_l} \ ,
\ee
so that by (\ref{homrelbis})
\be\label{f0f0hkgen}
\cF_0(S_{0,r},S_{l,p})\Big\vert_{S_{0,r}=0,\ S_{l,p}={1\over k}\Sh_l}
={1\over k} \cFh_0 (\Sh_l) \ .
\ee
In general, however, we do not have explicit expressions for 
$\cFh_0(\Sh_l)$. 

One can similarly relate the effective superpotentials 
for appropriate $N_{l,q}$, as before, and even show that 
$\Weff$ is symmetric under simultaneous 
cyclic symmetries $S_{l,q}\to S_{l,q+1}$, but as for the general 
even superpotential, this is not enough to determine any vacua.

\subsection{Superpotentials of the form $W(x)={1\over k} \Wh(h(x))$}

Finally, we consider the case of general mappings
\be\label{hfunction}
\xi=h(x) = x^k +h_{k-2} x^{k-2} + \ldots + h_1 x \ ,\qquad k\ge 3 \ ,
\ee
and superpotentials $W(x)={1\over k} \Wh(\xi)$, where $\Wh$ is of 
order $m+1$ in $\xi$. Note that we have set $h_{k-1}$ and $h_0$ to 
zero by appropriate shifts of $\xi$ and $x$. In particular, a quadratic map 
$h(x)=x^2+h_1 x +h_0$ can always be reduced to the case studied in 
section 3.1. Since $W'(x)={h'(x)\over k}\Wh'(\xi)$, we choose 
$f(x)=\left({h'(x)\over k}\right)^2 \wh f(\xi)$ 
so that
\be\label{y2yh2relh}
y^2(x)=\left({h'(x)\over k}\right)^2 \left[ \Wh'(\xi)^2+\wh f(\xi)\right] 
\equiv \left({h'(x)\over k}\right)^2 \yh^2(\xi) \ ,
\ee
and thus
\be\label{yyhrelh}
y(x)\d x ={1\over k} \yh(\xi)\d\xi \ ,
\ee
as before. 

Again, to each cut ${\cal C}_l,\ l=1,\ldots m$ of $\yh$ correspond 
$k$ (non-degenerate) cuts ${\cal C}_{l,q}$, $q=1,\ldots k$ of $y$, 
and the cycles 
$A_{l,q}$ are mapped to $A_l$. There are also $k-1$ degenerate cuts 
at the zeros of $h'(x)$. The picture is still as in Fig.\ \ref{x3-10}, 
but now the $\Z_k$-symmetry gets distorted. Obviously, eq. (\ref{sshkgen}) 
continues to hold: $S_{l,p}={1\over k} \Sh_l$ and, in particular, $t=\wh t$,
while $S_{0,r}=0,\ r=1,\ldots k-1$, corresponding to the double zeros of $y^2$.
For the $B_{l,q}$ cycles one proceeds as follows. First, one chooses $\L_0$
and defines $\Lh_0=h(\L_0)$. We call $\L_0^{(q)}$ the $k$ roots of 
$h(x)=\Lh_0$, labelled such that 
$\L_0^{(q)} \simeq \o^{q-1} \L_0 + {\cal O}\left({1\over \L_0}\right)$.
The $\wt B_{l,q}$ cycles then go from ${\L_0^{(q)}}'$ on the lower sheet 
through the cut ${\cal C}_{l,q}$ to $\L_0^{(q)}$ on the upper sheet 
and are mapped exactly, via $\xi=h(x)$, to the $B_l$ cycles which go 
from $\Lh_0'$ through ${\cal C}_l$ to $\Lh_0$.
Furthermore, defining the $B_{l,q}$ cycles appropriately, they can be 
decomposed into various $C_{\pm,r}$, $A_{l',q'}$ and the $\wt B_{l,q}$ 
cycles such that $\int_{B_{l,q}} y(x)\d x=\int_{\wt B_{l,q}} y(x)\d x
={1\over k}\int_{B_l}\yh(\xi)\d\xi$, as before. Since we still have
$W(\L_0)={1\over k}\Wh(\Lh_0)$ and 
$\log\L_0^2={1\over k}\log\Lh_0^2 + {\cal O}\left({1\over \L_0^2}\right)$
we conclude again that
${\del\cF_0\over \del S_{l,q}}={1\over k} {\del \cFh_0\over \del\Sh_l}$ 
at $S_{l,q}={1\over k} \Sh_l$, \ $S_{0,s}=0$ and, hence
$\cF_0(S_{0,s}=0;\,S_{l,q}={1\over k}\Sh_l)
={1\over k} \cFh_0 (\Sh_l)$, as before. Also, the relation between the 
effective superpotentials continues to hold, provided one makes the 
symmetric choice of the $N_{l,q}$.

If we specialise to the case $m=1$ where
$\Wh$ is a gaussian superpotential, i.e. for a
\be\label{whgauss}
W(x)={1\over 2k}h(x)^2 -{a\over k} h(x) +b \ ,
\ee
we know, of course, the exact expression of $\cFh_0(t)$.
In this case, one might ask further whether one can
still prove some permutation symmetry of $\Weff$, for unequal $S_i$, 
and use this to find vacua. However,
above, we exploited the $\Z_k$-symmetry of $W(x)$  to prove the symmetry 
under circular permutations of the $S_i$, and it seems unlikely that one 
can proceed without it.

\section{\bf Conclusions}

\vskip-2.mm
\begin{table}[h]
\begin{center}
\begin{tabular}{|c|c|c|c|c|c|}\hline
& & & & & \\
{}& Sect. 3.1 & Sect. 3.2 & Sect. 3.3 & Sect. 3.4 & Sect. 3.5 \\
$l=1,\ldots m$ &  $m=1$ &  $m\ge 2$ & $m=1$ & $m\ge 2$ & $m\ge 1$ \\
$r=1,\ldots k$, \ 
$s=1,\ldots k-1$ & $k=2$ & $k=2$ & $k\ge 3$ & $k\ge 3$ & $k\ge 3$ \\
& & & & & \\
\hline
& & & & & \\
{\rm map} 
& $\xi=x^2$
& $\xi=x^2$
& $\xi=x^k$
& $\xi=x^k$
& $\xi=h(x)$
\\
& & & & & \\
\hline
& & & & & \\
$\cF_0(S_{0,s}=0;\,S_{l,r}={\Sh_l\over k}) ={1\over k} \cFh_0(\Sh_l)$ &
yes & yes & yes & yes & yes \\
& & & & & \\
\hline
& & & & & \\
$\Weff(S_{0,s}=0;\,S_{l,r}
={\Sh_l\over k})$ & & & & & \\
$={1\over k} \Wheff(N_l;\Sh_l)$ &
yes & yes & yes & yes & yes \\
at $N_{l,r}={N_l\over k},\ N_{0,s}=0$ & & & & & \\
& & & & & \\
\hline
& & & & & \\
$\cF_0(S_{0,s}=0;\,S_{l,1},S_{l,2},\ldots S_{l,k})$ & & & & & \\
$=\cF_0(S_{0,s}=0;\,S_{l,k},S_{l,1},\ldots S_{l,k-1})$ &
yes & yes & no & no & no \\
& & & & & \\
\hline
& & & & & \\
$\Weff(S_{0,s}=0;\,S_{l,1},S_{l,2},\ldots S_{l,k})$ & & & & & \\
$=\Weff(S_{0,s}=0;\,S_{l,k},S_{l,1},\ldots S_{l,k-1})$ &
yes & yes & yes & yes & no \\
at $N_{l,r}={N_l\over k},\ N_{0,s}=0$ & & & & & \\
& & & & & \\
\hline
& & & & & \\
all vacua of $\Wheff$ yield vacua of $\Weff$ &
yes & ? & yes & ? & ? \\
& & & & & \\
\hline
\end{tabular}
\end{center}
\caption{\it The table summarises our results for the different 
pairs of superpotentials.} 
\label{table1}
\end{table}
\vskip 2.mm

In this note we studied relations between
effective superpotentials (as well as prepotentials)
of $\cN=1$ $U(N)$ gauge theories with different tree-level 
superpotentials $W$ and $\Wh$ for an adjoint chiral multiplet,
with particular emphasis on $W$'s that preserve an anomaly-free 
$\Z_k$ symmetry $\Phi\to e^{2\pi i/k}\Phi$.
These tree-level superpotentials which are polynomials of order 
$k(m+1)$ and  $m+1$, respectively, are related by $W(x)=\Wh(\xi(x))$. 
The determination of the effective superpotentials is essentially 
reduced to the computation of various period integrals on 
corresponding Riemann surfaces $\cR$ and $\cRh$, and $\xi(x)$ constitutes a 
map between them. For a ``general'' degree $k$ polynomial $\xi(x)$, 
this mapping provides the
relation between the effective superpotentials of the gauge theories, 
but also between the prepotentials or the free energies $\cF_0$ and $\cFh_0$ 
of the corresponding holomorphic 
matrix models in the planar limit. On the ``symmetric'' submanifold of
moduli space given by $S_{l,r}={1\over k} \Sh_l$ and $S_{0,s}=0$ 
we could express $\cF_0$ and 
$\Weff$ entirely in terms of $\cFh_0$ and $\Wheff$. Moreover, 
in the $\Z_k$ symmetric case $\xi=x^k$, for
unequal $S_{l,r}$, we could prove, for $k=2$, symmetry of $\cF_0$ 
under exchange
of the arguments $S_{l,1}\leftrightarrow S_{l,2}$ which is the 
quantum manifestation of the $\Z_2$ symmetry in this case. For 
$k\ge 3$ the $\Z_k$ symmetry does not completely 
survive at the quantum level 
and the corresponding permutation symmetry $S_{l,r}\to S_{l,r+1}$ of $\cF_0$
is anomalous due to subtleties in the precise definition of the
non-compact period integrals and the necessity to introduce a common 
``cut-off'' $\L_0$ for all of them. However, the anomalous term
is irrelevant when 
looking only at the effective superpotential for symmetric gauge group
breaking patterns, i.e. $N_{l,r}={1\over k} N_l$ 
and $N_{0,s}=0$, and the 
permutation symmetry is restored for all $k$. This in turn allowed us to show,
for $m=1$, that for each vacuum of $\Wheff$ (which in this case is 
the Veneziano-Yankielowicz superpotential) there is a 
corresponding vacuum of $\Weff$. All this is summarised in Table 1.

\end{document}